\documentclass[sigplan,review=false]{acmart}

\copyrightyear{2018}
\acmYear{2018}
\setcopyright{acmcopyright}
\startPage{1}
\setcopyright{none}

\usepackage{booktabs}   %% For formal tables:
\usepackage{subcaption} %% For complex figures with subfigures/subcaptions
\usepackage{enumitem}
\usepackage{listings}
\usepackage{array}
\usepackage{mathtools}
\DeclarePairedDelimiter{\ceil}{\lceil}{\rceil}

\usepackage{multirow}
\usepackage{ifthen}
\usepackage{nicefrac}
\usepackage{epigraph}

\usepackage[small,compact]{titlesec}
\usepackage[font={small}]{caption}  % small fonts in captions

\selectfont

\newboolean{showcomments}
\setboolean{showcomments}{true}
\ifthenelse{\boolean{showcomments}}
{ \newcommand{\mynote}[3]{
    \fbox{\bfseries\sffamily\scriptsize#1}
    {\small$\blacktriangleright$\textsf{\emph{\color{#3}{#2}}}$\blacktriangleleft$}}}
{ \newcommand{\mynote}[3]{}}

\newcommand{\stimes}{{\times}}

\newcommand{\FFTs}{\mathfrak{F}}
\newcommand{\FFTGs}{\mathfrak{G}}
\newcommand{\WINs}{\text{F}}

\newcommand{\AI}{AI}
\newcommand{\FPO}{FPO}
\newcommand{\DM}{DM}
\newcommand{\CMR}{CMR}
\newcommand{\MB}{MB}
\newcommand{\FLOPS}{Peak~FLOPS}

\newcommand{\remove}[1]{}

\newcommand{\Conv}{\mathop{\scalebox{1.5}{\raisebox{-0.2ex}{$\ast$}}}}%

\begin{document}

\setlength{\pdfpageheight}{\paperheight}
\setlength{\pdfpagewidth}{\paperwidth}

\title{FFT Convolutions are Faster than Winograd on Modern CPUs, Here's Why}

%\author{ \normalsize \bf Submission 4 }

%\remove{

\author{Aleksandar Zlateski}
\authornote{Both authors contributed equally to the paper}
\affiliation{ \institution{Massachusetts Institute of Technology} }

\author{Zhen Jia}
\authornotemark[1]
\affiliation{  \institution{Princeton University} }

\author{Kai Li}
\affiliation{  \institution{Princeton University} }

\author{Fredo Durand}
\affiliation{  \institution{Massachusetts Institute of Technology} }

%} %% end remove

\settopmatter{printacmref=false} % Removes citation information below abstract
\renewcommand\footnotetextcopyrightpermission[1]{} % removes footnote with conference information in first column
\pagestyle{plain} % removes running headers

\date{}
\thispagestyle{empty}

\begin{abstract}
%\zhen{Refine title}

Winograd-based convolution has quickly gained traction as a preferred approach to implement convolutional neural networks (ConvNet) on various hardware platforms because it requires fewer floating point operations than FFT-based  or direct convolutions.     
 \begin{sloppypar}
This paper compares three highly optimized implementations (regular FFT--, Gauss--FFT--, and Winograd--based convolutions) on modern multi-- and many--core CPUs. 
Although all three implementations employed the same optimizations for modern CPUs, our experimental results with two popular ConvNets (VGG and AlexNet) show that the FFT--based implementations generally outperform the Winograd--based approach,  contrary to the popular belief.
\end{sloppypar}

To understand the results, we use a Roofline performance model to analyze the three implementations in detail, by 
looking at each of their computation phases and by considering
not only the number of floating point operations, but also the memory bandwidth and the cache sizes.
The performance analysis explains why, and under what conditions, the FFT--based implementations outperform the Winograd--based one, on modern CPUs. 

%We hope that both experimental findings and performance analysis can help DNN library developers make better future design and implementation decisions.

\end{abstract}

\maketitle

%\setlength{\belowcaptionskip}{-10pt}
%\setlength{\abovecaptionskip}{5pt}

%\aleks{redo figure 4, so it deosn't say "our winograd" but rather winograd of Jia at al.}

\section{Introduction}
 
  Convolutional neural networks (ConvNets) have emerged as a
  widespread machine learning method for many application domains,
  soon after the demonstration of its superior performance for
  classification and localization tasks in the
  ImageNet\cite{imagenet_cvpr09} competition in
  2012~\cite{krizhevsky2012imagenet}.  Since convolutional layers are
  computationally intensive and dominate the total execution time of
  modern deep ConvNets~\cite{szegedy2015going, krizhevsky2012imagenet,
    montufar2014number, simonyan2014very}, many efforts have been made
  to improve the performance of the convolutional primitives for
  CPUs~\cite{zlateski2016znn, vanhoucke2011improving, budden2016deep,
    falconlib, zlateski2017compile}, GPUs~\cite{chetlur2014cudnn,
    mathieu2013fast, vasilache2014fast, neon} or both
  \cite{zlateski2016znni}.

  An important class of optimization is to reduce the number of
  calculations required for a convolution.  Initially, several efforts
  used FFT--based convolutions to reduce the required computations for
  GPUs~\cite{mathieu2013fast, vasilache2014fast} and
  CPUs~\cite{zlateski2016znn, zlateski2016znni}.  Recently Lavin et
  al.~\cite{lavin2016fast} proposed a Winograd--based method for
  performing convolutions, and demonstrated that it can save more
  floating point operations than FFT, especially for small 2D kernels (e.g. $3 \times
  3$).  This prompted a large number of implementations to employ 
  Winograd--based convolutions.  For example, Nervana~\cite{neon} and Nvidia's
  cuDNN~\cite{chetlur2014cudnn} implemented Winograd--based
  convolution for GPUs.  
  FALCON~\cite{falconlib}, LIBXSMM\cite{libxsmmGit} and Intel
  MKL-DNN~\cite{mkl-dnn} provided CPU implementations of Winograd-based
  convolutions.  Budden et al.~\cite{budden2016deep} 
  extended the 2D Winograd-based convolution to arbitrary dimensions and kernel sizes.  
  A highly optimized implementation for
  modern many--core CPUs was presented by~\cite{winograd}. However, the community has not put in similar optimizing efforts into FFT--based implementations.

  This paper presents results of comparing Winograd--based with FFT--based convolutions 
  on modern CPUs.  We have extended the highly optimized Winograd--based implementation for 
    Intel Xeon Phi processors~\cite{winograd}, to support arbitrary modern multi-- and many--core
    CPUs (that support the AVX2 or the AVX512 instruction set).
    Using the same optimization techniques, we additionally
    implemented two FFT--based algorithms for modern CPUs.  
    One is based on the regular FFT algorithm (Regular--FFT).  
    Another is also FFT--based but uses Gauss' multiplication method (Gauss-FFT).  
    All implementations are open--sourced at~\cite{deleted}.
    
    We have compared all three implementations with two popular ConvNets (VGG and AlexNet) on 10 different systems with modern CPUs.  
    Our results show that, contrary to the popular belief, the regular--FFT or Gauss--FFT implementations are generally faster than the Winograd implementation. And in some cases, they are substantially faster.
  
    To understand the experimental results, we have used the Roofline performance model~\cite{williams2008roofline} to analyze each
    algorithm and its implementation in detail by carefully analyzing its computational phases.  Such
    analysis provided two key explanations.  First, Winograd--based
    approach, in most cases, requires fewer floating point operations than FFT--based approach 
    because it works with real numbers instead of complex numbers.
    However, due to its numerical instability, the Winograd method can 
    use only very small transform sizes~\cite{lavin2016fast,
    vincent2017improving, budden2016deep}.  The FFT--based convolutions 
    do not suffer from such instabilities, allowing for 
    arbitrary large tile sizes.  Large tile sizes allow the FFT--based
    approach to reduce a large number of redundant or unnecessary 
    computations.  Thus, in certain scenarios, the FFT--based method
    requires fewer operations than the Winograd--based one.  
    
  Second, our analysis considers not only the number of 
  floating point operations, but also the total amount of data movements (to and
  from memory) and their costs, the arithmetic intensity (operations per moved byte),
  the processor's speed, as well as the memory bandwidth.  We also
  analyze the effects of cache sizes, which determine the arithmetic
  intensity of both methods, and indirectly affect the running times.
  
  Large arithmetic intensity allows for better utilization of hardware systems whose
  compute--to--memory ratios are increasing over time, because of
  the imbalance of the evolution between processor speeds and memory bandwidths.  The speeds for computations typically improve much faster than memory bandwidths~\cite{wulf1995hitting}~\footnote{For instance, the $4.5$ TFLOPS Intel Knights Landing
  processor~\cite{jeffers2016intel} has a compute--to--memory ratio of
  $11$, whereas the latest Skylake Xeon and SkylakeX family of
  processors have reached the ratio in the range between $20$ to almost
  $40$.}.  This benefits the FFT--based method more, as its arithmetic intensity 
  is generally higher due to computing with complex numbers. 
  
  Our analysis suggests that whether the Winograd--based or
  a FFT--based approach is faster depend on the specific convolutional
  layer and the particular system it is executed on.  However, on
  average, the FFT--based approaches outperform the Winograd--based
  one with commonly used neural networks, with the margin
  increasing as the system's compute--to--memory ratio increases.
  
  The findings in this paper challenge the current belief  
  that Winograd--based convolutions are better in nearly all practical cases.
  
\section{Background} \label{sec:background}

  \subsection{Winograd-- and FFT-- Based Convolutions}
  
   \subsubsection*{Winograd -- Based Convolution}

  As recently illustrated by Lavin et al.~\cite{lavin2016fast},
  ``valid'' convolution of discrete signals, the main operation used
  in modern convolutional neural networks, can be performed using
  Winograd's minimal filtering algorithm \cite{winograd1980arithmetic}
  via
  {\small
    \begin{equation} \label{eq:winograd}
      f \Conv_{\texttt{Valid}} g = \mathbb{A}^T[(\mathbb{G}g) \odot (\mathbb{B}^Tf)]
    \end{equation}
  }

  Where $f$ and $g$ are 1D signals; $\odot$ represents element--wise
  multiplication; and $\mathbb{A}^T$, $\mathbb{B}^T$ and $\mathbb{G}$
  are special matrices, derived from Vandermonde matrices for
  Homogeneous Coordinate polynomials~\cite{vincent2017improving}.  By
  convention, Winograd convolutions have the matrices $\mathbb{A}^T$,
  $\mathbb{B}^T$ and $\mathbb{G}$ in real--space.  In ``valid''
  convolution the filter slides across every ``valid'' location in the
  filtered images -- such that the filter is fully contained inside
  the image.  When the size of the filter is $|g| = r$, $f
  \displaystyle{\scriptsize{\Conv_{\texttt{\scriptsize{Valid}}}}}g$
  will have a length of $|f| - |g| + 1 = m$, and we refer to the
  method above as Winograd convolution $\WINs(m,r)$.
  
    \subsubsection*{FFT -- Based Convolution}

  The convolution theorem states that a convolution
  can be performed using Fourier transforms via
  {\small
    \begin{equation}
      f \Conv_{\texttt{Circ}} g = \mathcal{F}^{-1} \big( \mathcal{F}(f) \cdot
      \mathcal{F}(g) \big)
    \end{equation}
  }

  Here, $\mathcal{F}$ and $\mathcal{F}^{-1}$ are Fourier and inverse
  Fourier transforms.  In the discrete case, $f$ and $g$ need to have
  the same number of elements, which can be accomplished by padding
  zeros to the shorter signal.

  Discrete Fourier transform (DFT) results in a circular (also known
  as cyclic) convolution.  The result of the ``valid'' convolution can
  be extracted from the last $|f| - |g| + 1 = m$ elements of the
  circular convolution.

  ``Valid'' convolution using discrete FFTs can also be regarded as a
  special case of the Eq~\ref{eq:winograd}, where the matrices
  $\mathbb{A}^T$, $\mathbb{B}^T$ and $\mathbb{G}$ are in complex space
  and are derived from Vandermonde matrices with polynomial points
  being the roots of unity~\cite{vincent2017improving}.
  $\mathbb{B}^T$ and $\mathbb{G}$ perform, implicitly zero--padded (to
  size of $m + r - 1 = |f|$), DFT transforms of $f$ and $g$, and
  $\mathbb{A}^T$ computes the last $m$ elements of the inverse DFT
  transform.  Using the FFT algorithm allows for efficient computation
  of matrix--vector products with matrices $\mathbb{A}^T$,
  $\mathbb{B}^T$ and $\mathbb{G}$.  We refer to this special case as
  Regular--FFT $\FFTs(m,r)$.

  \subsubsection*{Multi--Dimensional Convolution}

  Both Winograd and FFT convolutions can easily be extended to an
  arbitrary number of dimensions~\cite{budden2016deep}.
  $N$--dimensional convolution is performed via
  {\small
    \begin{equation}
      f \Conv_{\texttt{Valid}} g = \Big[ (g \stimes_{n=1}^{N}
        \mathbb{G}) \odot (f \stimes_{n=1}^{N} \mathbb{B})
        \Big]\stimes_{n=1}^{N} \mathbb{A}_n^T
    \end{equation}
  }

  Here, the operation $x \times_n^N \mathbb{Y}$ is short for $x
  \times_1 \times_2 \dots \times_n \mathbb{Y}$, where $\times_n$
  represents tensor--matrix mode--n multiplication as defined
  in~\cite{kolda2009tensor, budden2016deep}.  For the 2D case, $x
  \stimes_1 \mathbb{M} = \mathbb{M}x$, and $x \stimes_2 \mathbb{M} =
  x\mathbb{M}^T$.  The formula above reduces to
  {\small
    \begin{equation}
      f \Conv_{\texttt{Valid}} g = \mathbb{A}^T \Big[ (\mathbb{G}g\mathbb{G}^T) \odot (\mathbb{B}f\mathbb{B}^T)
        \Big] \mathbb{A}
    \end{equation}
  }

  Which is consistent with Lavin et al.~\cite{lavin2016fast}.

  \subsection{Winograd-- and FFT-- Based Convolution Layers} \label{subsec-WandFLayer}

  A convolutional layer transforms an input tuple of $C$ images into
  an output tuple of $C'$ images.  A batch of $B$ inputs yielding a
  batch of $B$ outputs is processed at the time via
  { \small
    \begin{equation}\label{eq:forward}
      I'_{b,c'} = \sum_{c=1}^C I_{b,c}\Conv W_{c'c}
    \end{equation}
  }

  Where $C$ and $C'$ denote the number of input/output images (also
  called channels or feature--maps).  $I_{b,c}$ and $I'_{b,c'}$ are
  the (arbitrary dimensional) input and output images of the $b$--th
  batch.  In total, $B \cdot C \cdot C'$ convolutions are performed.

  Using Winograd or Regular FFT convolution, the output images are
  computed via
  \begin{equation}  \small %\scriptsize
    \begin{aligned}
      I'_{b,c'} & = \sum_{c=1}^C \Big[ (W_{c,c'} \stimes_{n=1}^{N}
        \mathbb{G}_n) \odot (I _{b,c} \stimes_{n=1}^{N} \mathbb{B}_n)
        \Big]\stimes_{n=1}^{N} \mathbb{A}_n^T \\ & = \Big[
        \sum_{c=1}^C (W_{c,c'} \stimes_{n=1}^{N} \mathbb{G}_n) \odot
        (I_{b,c} \stimes_{n=1}^{N} \mathbb{B}_n)
        \Big]\stimes_{n=1}^{N} \mathbb{A}_n^T \\
    \end{aligned}
    \label{eqn-layer-comp}
  \end{equation}

  Note that we can have different sizes for matrices $\mathbb{A}_n$,
  $\mathbb{B}_n$ and $\mathbb{G}_n$ for each dimension.

  $\WINs_n(m_n,r_n)$ and $\FFTs_n(m_n,r_n)$ assume a particular size
  of $I$ ($m_n + r_n - 1$) and $I'$ ($m_n$) along the $n$--th
  dimension.  For larger image sizes, the convolution is performed
  using the overlap--add method (OLA)~\cite{rabiner1975theory}.  With
  OLA, the input images are divided into tiles with sizes of $m_n +
  r_n - 1$, and an overlap of $r_n - 1$ along the $n$--th dimension.
  Considering tiles at the same location from all the input images,
  tiles of size $m_n$ of the output images are computed using the
  formula above.

  The main savings in computation in both the Winograd and FFT methods
  comes from the fact that both the kernel transforms $(W_{c,c'}
  \times_{n=1}^{N} \mathbb{G}_n)$, and image (tiles) transforms
  $(I_{b,c} \times_{n=1}^{N} \mathbb{B}_n)$ can be precomputed and
  reused many times.  The computation is dominated by computing the
  dot products -- accumulation of element--wise product inside the
  square brackets in Eqn.~\ref{eqn-layer-comp}.  Computing all the dot
  products in Eqn.~\ref{eqn-layer-comp} is an equivalent problem to
  matrix multiplications, with real matrices for the case of Winograd
  and complex matrices for the case of Regular FFT convolution.

  \subsection{Gauss' Multiplication of Complex Numbers} \label{sub_sec_gauss}

  In the Regular FFT convolution, the computation is dominated by
  complex matrix multiplications, where each complex number pair
  multiplication requires 4 real multiplications, when computed
  directly, and 3 real multiplications when using Gauss'
  multiplication algorithm~\cite{maclaren1970art,lavin2016fast}.

  Using Gauss' multiplication algorithm, the product of two complex
  numbers $u_r+u_ii$ and $v_r+v_ii$ is computed by first computing
  $tmp_1 = v_r·(u_r+u_i)$, $tmp_2 = u_r·(v_i-v_r)$ and $tmp_3 =
  u_i·(v_r+v_i)$. The real part of the result is then equal to $tmp_1
  - tmp_3$ and the imaginary part to $tmp_1 + tmp_2$.  Similarly, an
  element--wise product of two complex tensors $U \odot V$ ($U = U_r +
  U_ii$ and $V = V_r + V_ii$) can be performed using three
  element--wise products of real--valued tensors.

  For the Regular--FFT convolutional layer, element--wise product of
  complex tensors representing the image (tile) transforms and kernel
  transforms are performed, and each tensor is reused many times
  (Eqn.~\ref{eqn-layer-comp}).  After performing a transform of an
  image tile, which yields a complex tensor $U = U_r + U_ii$, a
  real--valued tensor $U_r + U_i$ can be computed and stored alongside
  $U_r$ and $U_i$.  Similarly, tensors $V_i-V_r$ and $V_r+V_i$ can be
  computed during kernel transforms and stored alongside $V_r$ ($V_i$
  does not have to be stored).  Each element--wise products of complex
  tensors can then be replaced with three independent element--wise
  products of real--valued tensors.

  The resulting three real tensors are implicitly converted back to a
  single complex tensor during the computation of inverse transform
  ($\stimes_{n=1}^{N} \mathbb{A}_n^T$).

  Computing all the dot products in Eqn.~\ref{eqn-layer-comp} is then
  performed using three real--valued matrix multiplications instead of
  a single complex matrix multiplication, reducing the number of
  required operations by 25\%.

  We refer to the FFT method using Gauss' multiplication as
  Gauss--FFT ($\FFTGs(m,r)$)

\section{Implementations} \label{sec:implement}

   Both Winograd and FFT approaches perform computations in four distinct stages: input transform, 
   kernel transform, element--wise computation, and inverse transform.  
   The first two stages convert the images/kernels from a
   spatial domain into Winograd or FFT domain.  The third stage
   is equivalent to matrix multiplications.  The inverse transform stage
   converts the results back to the spatial domain.

   We extended the publicly available Winograd implementation from~\cite{wConv, winograd}, which was highly optimized for many--core CPUs, to support arbitrary AVX2 and AVX512 multi--core CPUs. We used it as a base for our FFT implementations.  We reused most of the code (90\%) in order to leverage 
    the same optimization methods.
     
  \paragraph{Optimizations} \label{sec::imp_opt}
  
  In order to achieve high utilization of the hardware, both Winograd 
  and FFT methods use the identical optimizations as proposed in~\cite{winograd},
  including 
  %but not limited to, 
  software prefetching, memory and cache blocking, using aligned vector data 
  access and interleaving memory access with computation.
  
  We adopted the data layout proposed in~\cite{winograd,
  jeffers2016intel, zlateski2017compile} for input images, kernels
  and output, where $16$ images are interleaved in memory for easy
  vectorization.  In~\cite{winograd} $16$ was used due to the
  size of the AVX512 vector register, we keep it to $16$ regardless
  of the vector register size, as $16$ is the cache--line width (16
  32--bit floats), to facilitate efficient utilization of the memory
  subsystem.

  For data hand--off between the four stages of the algorithm,
  streaming stores to main memory were used, since the data will not
  be used in near future.  This saves memory bandwidth and avoids
  cache pollution. 
  
  Overall, all three implementations achieved high utilization of
  the available hardware, as discussed in the following sections.
  
  \paragraph{Transforms}

  To perform transforms of the input images and kernels, as well as
  the output images, the implementation of~\cite{winograd} provides
  C++ codelets that perform transforms of $16$ tiles at the same time.  The
  codelets are created by generating Winograd transforms using
  Wincnn~\cite{wincnn}, after which a computation graph is created and
  optimized, yielding codelets that utilize AVX512 instructions to
  transform 16 tiles at the time.

  For the FFT--based implementations, the codelets were replaced by
  C++ codelets generated using ``genfft'' supplied with
  FFTW~\cite{frigo1998fftw}.  ``Genfft'' was modified so that it can
  generate codelets that perform implicitly zero--padded FFT
  transforms, as well as computing only a subset of elements of the
  inverse transform.  Multidimensional (forward) transforms were
  implemented by combining codelets performing implicitly zero--padded
  real--to--complex transforms along one dimension, and ones
  performing complex--to--complex transforms along other dimensions.
  Backward transforms combined complex--to--complex transform codelets
  and complex--to--real ones. Different from existing FFT--based 
  convolutions, which limit transform size to small prime 
  factors~\cite{zlateski2016znni} or only numbers that are 
  powers of two~\cite{mathieu2013fast,cudnn}, our implementations 
  support arbitrary sizes.  

\begin{sloppypar}

  \paragraph{Element--Wise Multiplications}

  For the element--wise stage, where matrix--matrix multiplications
  are performed, the implementation of~\cite{winograd} provides JIT
  routines for real--matrix multiplications optimized for AVX512
  instruction set.  Following the same principles of~\cite{winograd}
  we implemented JIT real--matrix multiplication routines optimized
  for the AVX2 instruction set, as well as complex--number
  multiplication routines for both AVX512 and AVX2 instruction sets,
  which are required for the Regular--FFT method.
  
\end{sloppypar}

  \paragraph{Parallelization Through Static Scheduling}

  Each of the stages of our algorithm is parallelized using static
  scheduling originally proposed in~\cite{zlateski2017compile}, using
  the generalized implementation provided by~\cite{winograd}. To
  achieve optimal performance, each core is assigned roughly the same
  amount of computation.  The work is then executed using a single
  fork--join routine.

\section{Performance Comparisons} \label{sec::experiment}

   \begin{table}
     \centering
     \scriptsize
     \caption{Machine configurations used for benchmarking.  The MB represents the theoretical
     peak memory bandwidth, and CMR represents the ratio between the available FLOPS and
     the memory bandwidth.} %
     \begin{tabular}{p{63pt} |p{10pt}p{18pt}p{9pt}p{20pt}p{24pt}p{8pt}}
       \toprule
       CPU & Cores & GFLOPS & AVX & Cache & MB{\tiny(GB/s)} & CMR \\
       \midrule
       Xeon Phi 7210 & 64 & 4506 & 512 & 512 KB & 409.6 & 11 \\
       i7-6950X & 10 & 960 & 2 & 1 MB & 68.3  & 14.06 \\
       i9-7900X & 10 & 2122 & 512 & 1 MB & 96  & 22 \\
       Xeon Gold 6148 & 20 & 3072 & 512 & 1 MB & 128 & 24 \\
       E7-8890v3 & 18 & 1440 & 2 & 256 KB & 51.2 & 28.13 \\
       Xeon Platinum 8124M & 18 & 3456 & 512 & 1MB & 115.2 & 30 \\
       i9-7900X & 10 & 2122 & 512 & 1 MB & 68.3  & 31 \\
       Xeon Phi 7210 & 48 & 4506 & 512 & 512 KB & 102.4 & 33 \\
       Xeon Phi 7210 & 64 & 4506 & 512 & 512 KB & 102.4  & 39.11 \\
       i9-7900X & 10 & 2122 & 512 & 1 MB & 51.2  & 41.25 \\
       \bottomrule
     \end{tabular}
     \label{table:machines}
   \end{table}

  \begin{figure*}[ht]
    \begin{center}
      \includegraphics[width=1\linewidth]{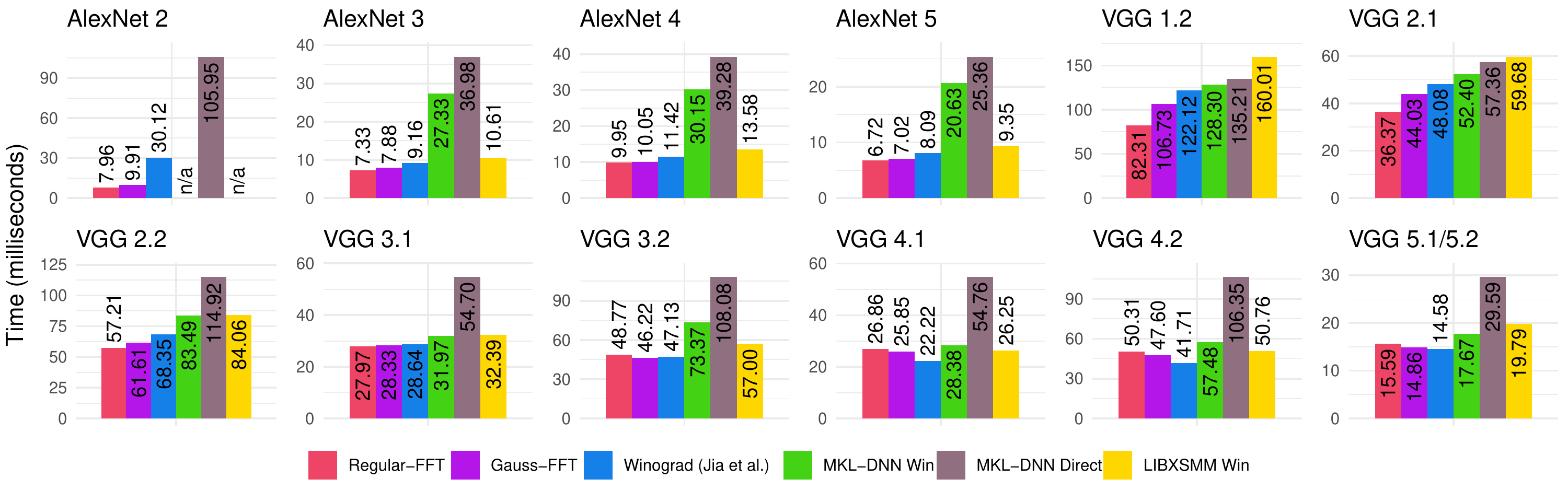}
    \end{center}
    \caption{Convolution layers' running time on Xeon Gold CPU.}
    \label{fig:combined-2d}
  \end{figure*}

  To compare running times of the FFT and Winograd methods,
  we benchmarked our FFT--based implementations and the improved Winograd 
  implementation of Jia et al.~\cite{winograd} by using the two most
  popular ConvNets, AlexNet~\cite{krizhevsky2012imagenet} and VGG~\cite{simonyan2014very},
  which are frequently used for benchmarking~\cite{imagenetBench}.  
  We benchmarked the time required for the forward propagation of each 
  distinct layers of the two networks.
  
  Additional, publicly available, implementations were included for reference --
  The Winograd implementations provided by LIBXSMM~\cite{libxsmmGit} and MKL-DNN~\cite{mkl-dnn}, and the direct convolution provided by 
  MKL-DNN~\cite{mkl-dnn}.  To the best of our knowledge, no
  other implementation of FFT--based methods for CPUs, beside
  our own, is currently available.  
  
  Both Winograd and FFT methods can work with arbitrary transform sizes.
  Generally, larger transform size decrease the number of required
  operations.  However, the numerical inaccuracy of Winograd 
  convolutions increases exponentially with transform (tile) sizes~\cite{pan2016bad,
    winograd, lavin2016fast}.  
  In practice, there is an upper bound on the transform size for which 
  the Winograd produces satisfactory results.  
  All major vendors, such as FALCON, MKL-DNN, LIBXSMM,
  and cuDNN~\cite{falconlib,libxsmmGit,mkl-dnn,cudnn} implement the
  Winograd algorithm with transform sizes less than or equal to $6\times 6$
  \footnote{With the transform sizes of $6 \times 6$, the average numerical error
  of the Winograd method on benchmarked layers was $7.03 \cdot 10^{-6}$, which is 
  similar to the error of direct--convolution $1.11 \cdot 10^{-6}$
  When the transform size
  is increased to $8 \times 8$, the error is increased to $1.24 \cdot 10^{-3}$, which was expected~\cite{winograd}.
  The numerical error of the FFT--based method was no larger than $2.88 \cdot 10^{-7}$, 
  regardless of the tile size.}.
  For these tile sizes, the numerical error of computation is 
  comparable to the one that implements 
  convolution directly. 
    
  We follow the same convention, and consider Winograd convolutions only with
  tile sizes of at most $6 \times 6$.  However, we allow the FFT methods to
  have an arbitrary large tile size, as they don't suffer from such numerical instability.

  \paragraph{Running Time Experiments}
  The benchmarks were performed on a total of 
  10 systems showed in Tbl.~\ref{table:machines}.  

  In Fig.~\ref{fig:combined-2d} we show detailed results on one of the systems.  
  Note that both LIBXSMM's and MKL-DNN's Winograd implementation support only
  kernel sizes of $3\stimes 3$, and thus can not use the Winograd method for the second layer of AlexNet. 
  
  The Winograd--based 
  method outperformed in only 3 out of 12 layers, whereas the a FFT--based
  method outperformed on 6 layers; and on 3 layers, they had roughly the 
  same performances.  More importantly, in the cases when the
  FFT methods outperformed, they did it with a larger margin, and
  on the layers that require more computation time.  This suggests
  that the FFT methods can allow for significant savings in the
  overall computation time, when compared to the Winograd.  
  For instance, the time spent on all convolutional layers
  in AlexNet using the Winograd method would consume 58.79 milliseconds, whereas the Regural--FFT method requires only 31.96 milliseconds; 
  that is a speedup of 1.84x.
  
\begin{figure}[ht]
    \begin{center}
      \includegraphics[width=1\linewidth]{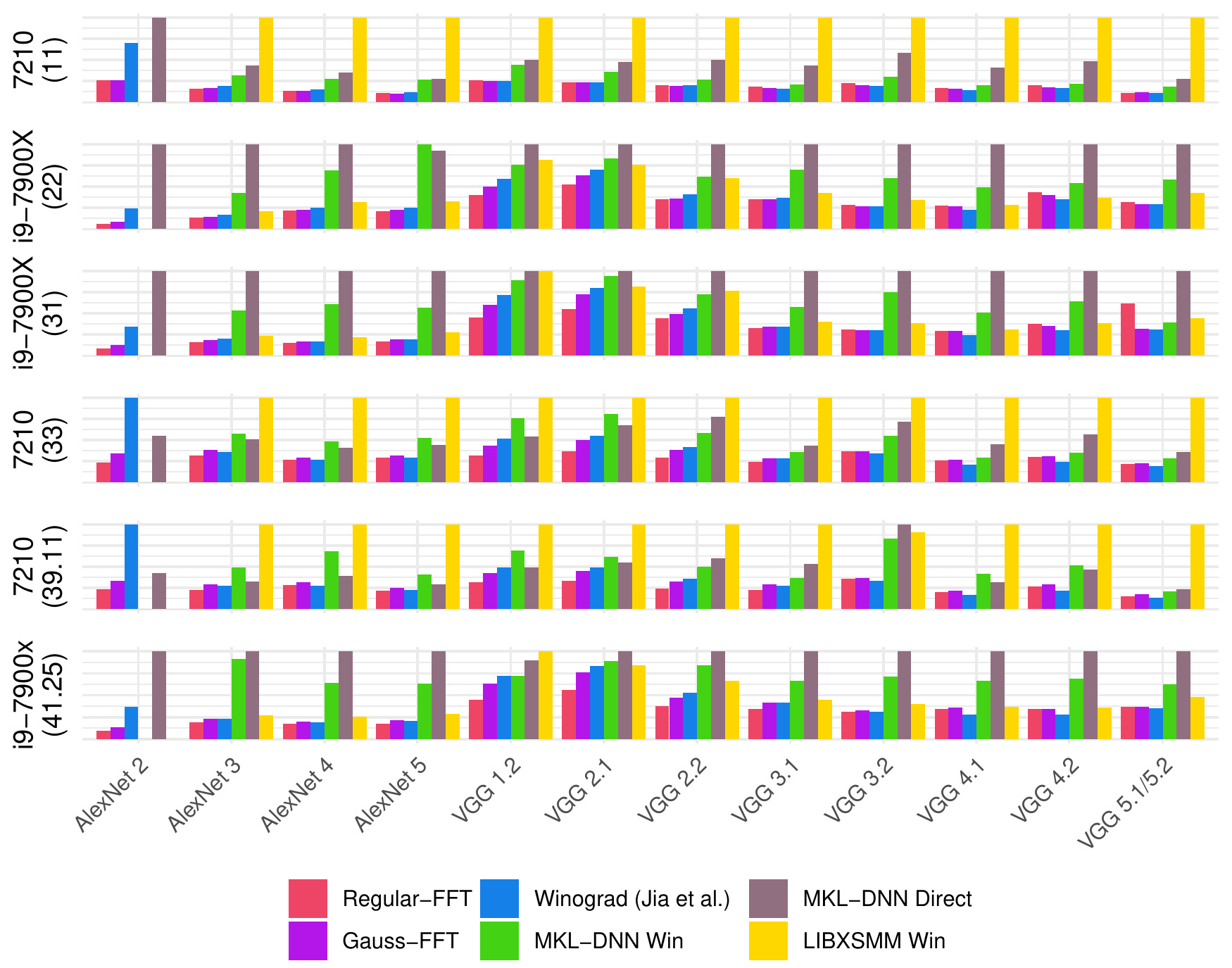}
    \end{center}
    \caption{Normalized running time of different convolution implementations. 
    (The systems with the same CPUs are distinguished by their \CMR~).}
    \label{fig:combined-2dextra}
  \end{figure}

  Additionally, in Fig.~\ref{fig:combined-2dextra} we show the normalized running time on
  all other AVX512 systems (neither LIBXSMM, nor MKL-DNN support the AVX2 instruction set).
  The results for each individual layer are scaled based on the slowest implementation,
  as we are only interesting in the relative performances. Detailed results are available in Appendix.
  In all cases, our two FFT--based implementations
  as well as the modified Winograd implementation of~\cite{winograd} outperformed other
  publicly available implementations.
  Thus, for the rest of the paper, we will only focus on these three implementations.
  
  \paragraph{FFT transform sizes} An important observation was that the 
  optimal transform sizes
  for the FFT method were not always powers of two;  they were
  27 for VGG1.2, 25 for VGG2.1 and VGG2.2, 21 for VGG 3.1 and VGG3.2,
  16 for VGG4.1 and VGG4.2, and 9 for VGG5.1/5.2.  For AlexNet,
  the optimal sizes were 31 for the second layer, and 15 for all other layers.
  This is counter intuitive, as the common belief is that
  optimal FFT transforms should use sizes that only contain
  factors that are small primes~\cite{frigo1998fftw,zlateski2016znni}
  or transforms with sizes equal to the powers of two, which is suggested by
  the two GPU FFT--based implementations, i.e., 
  fbfft~\cite{mathieu2013fast, vasilache2014fast} and CuDNN~\cite{chetlur2014cudnn}.

  \begin{sloppypar}
  \paragraph{Relative performances}
  While an FFT--method outperformed the Winograd more often than not, the relative performances varied among different layers and systems.  In the rest of the paper
  we focus on the theoretical analysis of all the methods.  We would like to
  understand why our findings are not aligned with the popular belief that the
  Winograd method should be strictly faster.
  \end{sloppypar}

\section{Performance Analysis} \label{Sec_analysis}

  Our experimental observations suggested that some of 
  the stages in both Winograd-- and FFT--based approach have relatively low
  utilization of the system's available FLOPS.  In most, but not all, cases, 
  these were the transform stages, which have relatively small amount of compute,
  but access relatively large amount of data, suggesting that their running time
  is bound by the memory bandwidth, and not the computational capabilities of
  the CPU.
  
  For this reason, we used the Roofline~\cite{williams2008roofline} performance model to analyze 
  Winograd-- and FFT-- based convolutions.

  \paragraph{Roofline Performance Model}  
  is an ISA oblivious, throughput
  oriented, performance model used to estimate performance of an application 
  running on processing units, and is often used to predict performances on
  CPUs, GPUs, TPUs (Tensor Processing Units), etc. 
  
  It is suitable for analyzing particular methods on systems where the 
  memory bandwidth often becomes the constraining resource. 
  It estimates the performance bound of an algorithm as a function
  of its arithmetic intensity (\AI), which is defined as the ratio of
  total floating--point operations (\FPO) and the total data movement
  (\DM) in bytes (\AI = \FPO/\DM) between different levels of the memory hierarchy.  
  For each level of memory hierarchy, the performance ceiling (attainable
  FLOPS) is determined by:
  \begin{equation}
    \label{eq:roofline}
    \small
    \begin{aligned}
      &\text{Attainable~ FLOPS} = \min(\FLOPS, ~\MB \times  \AI)
    \end{aligned}
  \end{equation}

  Where \FLOPS~ is defined as the maximal number of floating 
  operations per second of a given system, 
  and \MB~ as system's peak memory bandwidth.  
  When plotted, the performance ceiling line resembles a roofline.

  Here, we are interesting in the \DM~ between the 
  highest level of on--chip, core--exclusive cache (typically L2 for
  CPUs) and off--chip, shared memory~\footnote{The term off--chip
    refers to the main memory of the system, typically large, but much
    lower throughput and higher latency than the on--chip caches.  The
    HBW MCDRAM of the Knights Landing processor would be considered
    off--chip.}.  In a systems where the L2 is shared among a small
  number of cores (e.g. Intel Xeon Phi series share L2 cache between
  two cores.), the L2 is assumed to be equally divided for exclusive
  usage among the cores.

  \DM~ then accounts for all of regular and streaming stores to main
  memory, regular loads from main memory, and pre--fetches from main
  memory to either L1 or L2.

  Our analysis does not consider the presence of higher level, shared
  caches, as on modern processors, the sizes of such caches are very
  limited, and can not fit the working set of a typical convolutional layers.

  Additional performance ceilings for lower levels of cache, are not 
  necessary, since in both
  Winograd--based and FFT--based convolutional algorithms the computation 
  is either bounded by transfers to and from
  main memory, or the processor peak FLOPS, as shown in
  the following sections.

  As the performance ceiling is set by algorithm's arithmetic
  intensity (Eqn.~\ref{eq:roofline}), its running time can be
  estimated by:
  \begin{equation}
    \label{eq:running time}
    \footnotesize
    \begin{aligned}
      running~time & =  \frac{\FPO}{\text{Attainable~FLOPS}} \\
      & =  \frac{\FPO}{\min(\FLOPS, ~\mathsf{\MB} \times \AI )} = \frac{\FPO/\MB}{\min(\CMR,\AI)} \\
      & =  \begin{cases}
        \frac{\FPO}{\mathsf{Peak~ FLOPS}} ,& \mathsf{\CMR}  <= \AI\\
        \\
        \frac{\DM}{\mathsf{\MB}}  ,& \mathsf{\CMR}  > \AI
      \end{cases}
    \end{aligned}
  \end{equation}
  Here, \CMR~ is the system's compute--to--memory ratio, defined as
  the ratio of it's \FLOPS~ and \MB~ -- the memory bandwidth.  \FPO~
  represents the total number of floating point operations required,
  and \DM ~(the total amount of data movements in bytes), as defined
  above.  The running time is \emph{compute bound} when $\CMR <= \AI$,
  in which case it can be computed as $\frac{\FPO}{\FLOPS}$;
  otherwise, the running time is \emph{memory bound}, and can be
  computed as $\frac{\DM}{\MB}$.

  \paragraph{Estimating running time}  For the Wingorad-- and FFT--based convolutional layers, where the computation is
  composed of several sequential stages ($S$), each
  with a unique arithmetic intensity (\AI), the running time is
  calculated by accumulating the running times of each stage $s \in
  S$:
  \begin{equation}
    \label{eq:running time_total}
    \small
    \begin{aligned}
      running~time_{Total} = \displaystyle\sum_{}^{S}running~time_{s}
      = \displaystyle\sum_{}^{S}
      \frac{FPO_s/\text{MB}}{min(\text{CMR}, AI_s)}
    \end{aligned}
  \end{equation}

  \subsection{Relative Performances}

  We are interested in the relative performance between Winograd
  and FFT methods.  We define $Speedup(\mathcal{A},\mathcal{B})$ as
  the ratio of the running times of an algorithm $\mathcal{B}$ and an
  algorithm $\mathcal{A}$.
  \begin{equation}\label{eq:ratio_total} \small %\scriptsize
    \begin{aligned}
      Speedup(\mathcal{A},\mathcal{B}) &=
      \frac{running~time^{\mathcal{B}}_{Total}}{running~time^{\mathcal{A}}_{Total}}
    \end{aligned}
  \end{equation}
  A speedup greater than one indicates that the algorithm
  $\mathcal{A}$ is faster, and smaller than one means that the
  algorithm $\mathcal{B}$ is faster.

  While Eqn.~\ref{eq:running time_total} estimates the running time of
  an algorithm assuming perfect utilization of the hardware, which is
  rarely possible in practice.  However, the Eqn.~\ref{eq:ratio_total}
  is also valid when the compared algorithms are equally optimized,
  meaning that they utilize the same percentage of the hardware
  capabilities.

  In the Eqn.~\ref{eq:ratio_total} the value of \AI~ will differ
  between Winograd and FFT based approaches, and will also depend on
  the amount of available cache size~\cite{winograd}.
  Therefore, when comparing performance of two algorithms,
  the relative performance on a given system will depend on the \CMR~
  ratio and the amount of available cache, but not on the absolute
  values of the system's compute speed or memory throughput.
  
  {\bf Detailed analysis on obtaining the values of \AI, \DM, and \FPO~ are
  presented in the Appendix}~\ref{appendix_analysis}.

  Using Eqn.~\ref{eq:ratio_total} and the values of \AI, \DM, and \FPO~
  calculated in the Appendix~\ref{appendix_analysis} (Tbl.~\ref{table:layer-stuff}), we estimate
  the speedup between the FFT--based methods and the Winograd--based one.
  For all three methods, the tile sizes are chosen, in a way that minimizes
  the total running time (Eqn.~\ref{eq:running time_total}).
  
  \subsection{The Accuracy of the Theoretical Estimates}
  
   \begin{figure*}[ht]
     \begin{center}
       \includegraphics[width=1\linewidth]{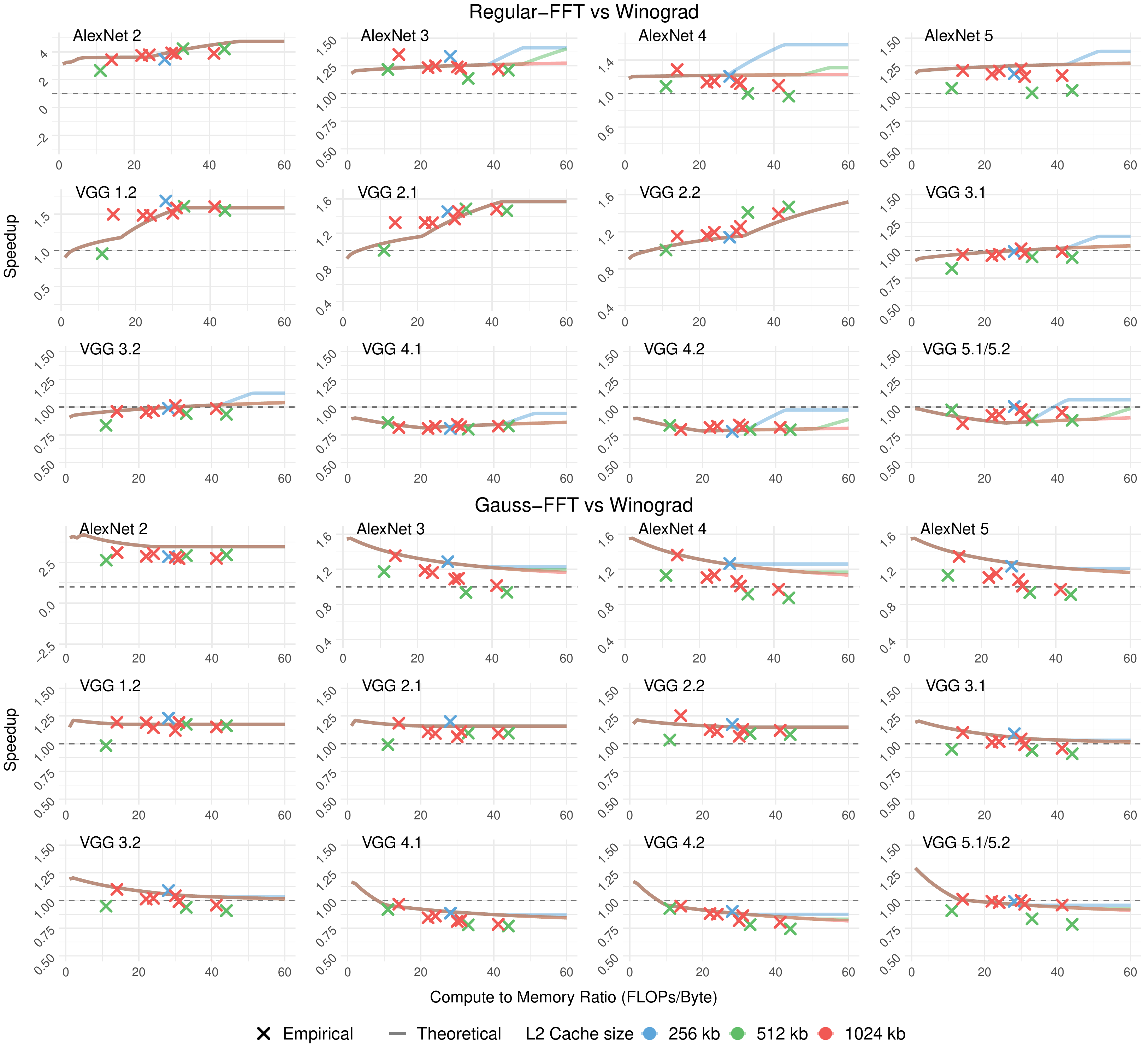}
     \end{center}
     \caption{Theoretical estimates and empirical measurements for the
       speedup of Regular-- and Gauss--FFT methods over the Winograd
       method on VGG and AlexNet.  The $x$ axis represent the system's \CMR.}
     \label{fig:model}
   \end{figure*}

  In Fig.~\ref{fig:model}, we plot the estimated theoretical speedup of the two FFT--based methods
  over the Winograd--based one.
  The solid lines represent the theoretical estimates of the relative performances 
  as a function of the system's \CMR.  The color of the line represents the amount of available
  L2 cache.  The lines are drawn in a semi--translucent fashion, as they overlap in many places.
  
  The empirical results from the benchmarks described in Section~\ref{sec::experiment} are
  overlaid; each cross--hair represents the measurement of
  the relative performances on a single system.  
  The $x$ coordinate of the cross-hairs represents the 
  system's compute--to--memory (CMR) ratio (see Tbl.~\ref{table:machines}), 
  and the color represents the L2 cache sizes.
  
  Our empirical results are consistent with our theoretical estimates. 
  The overall relative root mean square error (\textbf{rRMSE}) was 0.079 for 
  Regular--FFT vs Winograd and 0.1 for Gauss--FFT vs Winograd.  The
  fitness~\footnote{$\text{fitness} = \frac{100}{1 + rRMSE}$} was
  $92.68\%$ for Regular--FFT vs Winograd and $90\%$ for Gauss--FFT vs
  Winograd.

\subsection{Discussions}
 
  \paragraph{Hardware utilization}
 
  We have also measured the system utilization of each stage of the three methods.
  While the utilization varied across benchmarked layers and systems, on 
  average, during the compute bound stages, 75\% of theoretical peak FLOPS
  were attained;  in memory bound stages slightly more than 85\% of the 
  theoretical memory bandwidth was achieved.
  This resulted in a somehow lower \emph{effective} \CMR,
  which is consistent with the results shown in Fig.~\ref{fig:model}.
  The empirical results are slightly shifted to the left (towards
  lower values of \CMR), when compared to the theoretical predictions, 
  which assume equal utilization of both the FLOPS and the memory bandwidth.

  \paragraph{Optimality of Tile Transforms}

  Both ``wincnn''~\cite{wincnn} and ``genfft''~\cite{frigo1998fftw}
  use heuristics to minimize the number of operations for performing
  transforms, and might not be optimal.  However, as their \AI s are
  much smaller than the \CMR s of modern systems, the computation is
  memory bound, and the running time of the transform stages will
  depend solely on the amount of data movement.  The largest \AI~ of the
  FFT transforms is 5.55, and for Winograd 2.38, much lower than \CMR s of the 
  modern CPUs.  The Xeon Phi Knights Landing processor has the \CMR~ of 11 (due to on--chip fast MCDRAM), and
  the Xeon Server processor family has \CMR s in the range of 20--40.
  Hence, our theoretical analysis yield identical estimates as if the optimal
  number of operations were provided.

  This is consistent with our experimental findings, where in some cases, 
  tile sizes that were large primes, such as 31, were optimal.  
  Using such sizes,
  the images could be divided into overlapping tiles with minimal overhead (minimal padding
  of the original image), which reduces both the number of required operations, and the
  amount of required data movements.

\section{Conclusions}
  
  This paper presents experimental and analytical findings that
  FFT--based convolutions are, on average, faster than Winograd--based 
  convolutions on modern CPUs.
  
  In contrast to the popular belief that the Winograd--based method is 
  more efficient, our experimental results of a highly optimized
  Winograd--based implementation and two similarly optimized
  FFT-based implementations on modern CPUs 
  show that the FFT convolutions are, more often than not, faster than Wingorad ones 
  for most commonly used ConvNet layers. 
  
  Our analysis using the Roofline model shows that whether the 
  Winograd-- or FFT--based approach is faster depends on both the layer and target hardware. 
  However, with the tendency of increasing compute--to--memory ratio of
  systems with modern CPUs, the FFT--based methods tend to be faster than Winograd.
  
  While, we primarily focused on modern multi-- and many--core
  CPUs, which generally have larger caches, but smaller memory bandwidths than modern GPUs, 
  we believe that our performance analysis can be applied to GPUs.
  Future work might include implementing and 
  benchmarking efficient GPU--based implementations and validating performance analysis based on the Roofline model.

%\clearpage

\bibliography{znnfft}

%\clearpage
%\input{section/appendix}
\appendix

\section{Detailed Analysis} \label{appendix_analysis}

   %Winograd and FFT approaches execute in a similar fashion,
   %performing computations in four distinct stages: input transform, 
   %kernel transform, element--wise computation, and inverse transform.  
   %The first two stages convert the images/kernels from a
   %spatial domain into Winograd or FFT domain.  The third stage
   %is equivalent to matrix multiplications.  The inverse transform stage
   %converts the results back to the spatial domain.

  For simplicity, we assume 2D and isotropic images and kernels, as well as computation being performed with 32--bit floating point numbers (4 bytes per number).  
  Extension to non--isotropic and $N$--dimensional
  images and kernels, as well as lower precision floats, is straightforward.  The benchmarked implementations,
  described in Section~\ref{sec:implement} support an arbitrary number of
  dimensions, as well as non--isotropic kernels.

  \begin{sloppypar}
  We follow the same notation as in Section~\ref{sec:background}, and
  for an arbitrary layer, use Winograd convolution $\WINs(m^2, r^2)$,
  Regular--FFT convolution $\FFTs(m^2, r^2)$ and Gauss--FFT
  convolution $\FFTGs(m^2, r^2)$.  Here, $m$ can take an arbitrary
  value, whereas $r^2$ has to equal to the sizes of the kernels in the
  layer.
  \end{sloppypar}
  
  We proceed to present the details of all three methods and
  estimate the total number of operations and data movement required in 
  each stage, from which we also calculate the arithmetic
  intensity of each of the three method.

  \subsection{Input Transform Stage}

  In Winograd and both FFT approaches, each of the $B \cdot C$ images
  ($B$ batches, each with $C$ input channels) is divided into
  overlapping tiles, which are then transformed.  Each tile has the
  size of $t^2$ with $t = (m + r - 1)$, and there is an overlap of $r
  - 1$ pixels between adjacent tiles along both dimensions.

  Each image of size $x \times x$ will thus be divided into $N =
  \ceil{(x-r+1)/m}^2$ tiles.  If necessary, the image will be
  implicitly zero--padded.  This yields a total of
  $BC\ceil{(x-r+1)/m}^2 = BCN$ image tiles.

  \paragraph{Number of required operations}

  Each tile ($I_{b,c,n}$) is transformed via
  $\tilde{I}_{b,c,n} = \mathbb{B}I_{b,c,n}\mathbb{B}^T$, which is a
  composition of two matrix--matrix multiplications.  Here $I_{b,c,n}$
  is the $n$--th tile of the $c$--th channel in batch $b$, and
  $\tilde{I}_{b,c,n}$ is its transform.  Both $\mathbb{B}$ and
  $I_{b,c,n}$ have the size $t^2$, requiring a total of $4t^3$
  operations ($2t^3$ per matrix--matrix multiplication).  Operations
  are on real numbers for the Winograd approach, and on complex
  numbers for the FFT approaches.

  The transforms can, however, be performed with much fewer
  operations.  In the Winograd method, the matrices $\mathbb{B}$ are
  sparse, and contain pairs of columns with similar numbers, allowing
  for reuse of many intermediate results.  In the FFT methods, instead
  of matrix multiplications, 2D real--to--complex DFT transforms are
  performed, which require much fewer operations ($\mu n^2\log n$
  instead of $\mathcal{O}(n^3)$).  Here, the constant $\mu$ can vary
  significantly based on the size of the transform
  performed~\cite{frigo1998fftw}.  It takes small values, when the
  size of the transform is a power of two, and larger values when the
  transform size contains large prime factors.

  Instead of using a closed--form expression that gives bounds on the
  number of operations required, we counted the number of operations
  in real, optimized, implementations and stored them in a lookup tables.
  For Winograd, we used the Winograd matrix generator~\cite{wincnn}, 
  as proposed in~\cite{winograd, budden2016deep} and further reduced the 
  number of operations using the simple optimizer provided by~\cite{winograd}.  
  For the FFT methods, we used the FFT codelet generator ``genfft'' from the FFTW
  library~\cite{frigo1998fftw}.  We denote the number of operations required for
  transforming an image tile as $\WINs^{I}(m^2,r^2)$ for the Winograd,
  $\FFTs^{I}(m^2,r^2)$ for Regular--FFT, and $\FFTGs^{I}(m^2,r^2)$ for
  Gauss--FFT. The pre--computed lookup tables are included in our 
  open--sourced repository.

  The total number of operations required for performing all the
  transforms for all three methods are given in
  Tbl.~\ref{table:layer-stuff}.

  \begin{table*} \centering
    \small
    \setlength\tabcolsep{2.5pt}
    \caption{FLOPS, DM and AI for different stages of the Winograd
      ($\WINs(m^2,r^2)$), Regular--FFT ($\FFTs(m^2,r^2)$) and
      Gauss--FFT ($\FFTGs(m^2,r^2)$) methods.  $N$ is the number of
      tiles per single image, the tiles have size of $t^2$ ($t =
      m+r-1$). The values of $c$ and $c'$ depend on the amount of available cache, and
      are obtained by solving the optimization problem given in Eqn.~\ref{eq:gemm_AI}.}
    \begin{tabular}{cc | ccc}
      \toprule
    %  & Stage & $\WINs(m^2,r^2)$  & $\FFTs(m^2,r^2)$ & $\FFTGs(m^2,r^2)$ \\
      & Stage & Winograd  &  Regular--FFT & Gauss--FFT \\
      \midrule
      \multirow{3}{*}{\rotatebox{90}{\textbf{FLOPS}}}
      & Input image transform & $BCN\WINs^{I}(m^2,r^2)$ & $BCN\FFTs^{I}(m^2,r^2)$ & $BCN\FFTGs^{I}(m^2,r^2)$ \\
      & Kernel transform & $CC'\WINs^{K}(m^2,r^2)$ & $CC'\FFTs^{K}(m^2,r^2)$ & $CC'\FFTGs^{K}(m^2,r^2)$ \\
			& Element--wise Product&$2t^2BNCC'$ &$8t\ceil{(t+1)/2}BNCC'$ & $6t\ceil{(t+1)/2}BNCC'$ \\
      & Output transform & $BC'N\WINs^{O}(m^2,r^2)$ & $BC'N\FFTs^{O}(m^2,r^2)$ & $BC'N\FFTGs^{O}(m^2,r^2)$ \\
      \midrule
      \multirow{3}{*}{\rotatebox{90}{\textbf{DM}}}
      & Input image transform & $4BCx^2 + 4BCNt^2$ & $4BCx^2 + 8BCNt\ceil{(t+1)/2}$ & $4BCx^2 + 12BCNt\ceil{(t+1)/2}$ \\
      & Kernel transform & $4CC'\big (r^2 + t^2 \big)$ & $4CC' \big ( r^2 + 2t\ceil{(t+1)/2} \big )$ & $4CC' \big ( r^2 + 3t\ceil{(t+1)/2} \big )$ \\
			& Element--wise Product&$4t^2BN(c + \alpha c')\dfrac{CC'}{cc'}$ &  $8t\ceil{(t+1)/2}BN(c + \alpha c')\dfrac{CC'}{cc'}$ & $12t\ceil{(t+1)/2}BN(c + \alpha c')\dfrac{CC'}{cc'}$ \\
      & Output transform & $4BC'N \big ( t^2 + m^2 \big)$ & $4BC'N \big( 2t\ceil{(t+1)/2} + m^2 \big )$ & $4BC'N \big ( 3t\ceil{(t+1)/2} + m^2 \big )$ \\
      \midrule
      \multirow{3}{*}{\rotatebox{90}{\textbf{AI}}}
      & Input image transform & $\dfrac{N\WINs^{I}(m^2,r^2)}{4x^2 + 4Nt^2}$ & $\dfrac{N\FFTs^{I}(m^2,r^2)}{4x^2 + 8Nt\ceil{(t+1)/2}}$ & $\dfrac{N\FFTGs^{I}(m^2,r^2)}{4x^2 + 12Nt\ceil{(t+1)/2}}$ \\
      & Kernel transform & $\dfrac{\WINs^{K}(m^2,r^2)}{4r^2 + 4t^2}$ & $\dfrac{\FFTs^{K}(m^2,r^2)}{4r^2 + 8t\ceil{(t+1)/2}}$ & $\dfrac{\FFTGs^{K}(m^2,r^2)}{4r^2 + 12t\ceil{(t+1)/2}}$ \\
		& Element--wise Product& $\dfrac{cc'}{2(c + \alpha c')}$ &  $\dfrac{cc'}{c + \alpha c'}$ & $\dfrac{cc'}{2(c + \alpha c')}$ \\
      & Output transform & $\dfrac{\WINs^{O}(m^2,r^2)}{4t^2 + 4m^2}$ & $\dfrac{\FFTs^{O}(m^2,r^2)}{8t\ceil{(t+1)/2} + 4m^2}$ & $\dfrac{\FFTGs^{O}(m^2,r^2)}{12t\ceil{(t+1)/2} + 4m^2}$ \\
      \bottomrule
    \end{tabular}
    \label{table:layer-stuff}
  \end{table*}
  
  \paragraph{Data movement}

  The tile sizes are relatively small, much smaller than available
  cache; thus, once a tile is fetched from main memory, all the
  computation is done in--cache.  Additionally, the overlapping
  regions between tiles need to be fetched from memory only once, as
  they can be stored in cache and reused for adjacent tiles.

  The total data that has to be moved from main memory to cache is
  thus $BCx^2 \cdot 4$ bytes -- reading each of the $B \cdot C$ images
  of size $x^2$ only once.  Each of the $BCN$ transformed tiles is
  stored back to main memory.  The size for each transformed tile will
  depend on the method.

  In Winograd convolution, transformed tiles are real tensors, and
  require a single 32--bit float per element, for a total of $4t^2$
  bytes.

  The FFT methods perform 2D FFT transforms of each tile.  As the FFTs
  are performed on real tensors, the resulting transforms will be
  complex, conjugate--symmetric (along one of the dimensions). Only
  $t\ceil{(t+1)/2}$ complex numbers need to be stored.  The
  Regular--FFT requires two real numbers per complex number, for a total of
  $8t\ceil{(t+1)/2}$ bytes per tile, and the Gauss--FFT requires three
  reals per complex number, a total of $12t\ceil{(t+1)/2}$ bytes per
  tile.

  The total amount of data movement for the three methods, as well as
  their arithmetic intensities (\AI s), is shown in
  Tbl.~\ref{table:layer-stuff}

  \subsection{Kernel Transform Stage}

  In all methods, each of the $C \stimes C'$ kernels (with size $r
  \stimes r$) is transformed via $\tilde{W}_{cc'}=\mathbb{G}W_{cc'}\mathbb{G}^T$.  
  The matrix $\mathbb{G}$ has size $t \stimes r$.
  
  Computing the transform of a kernel is an equivalent procedure to
  performing the input transform of a zero--padded kernel, to match 
  the size of the input tiles.  In practice, the transforms are 
  computed with implicit zero--padding.  As in the input stage, 
  we have pre--computed the number of operations required for transforming
  a kernel in all three methods: $\WINs^{K}(m^2,r^2)$
  for the Winograd, $\FFTs^{K}(m^2,r^2)$ for Regular--FFT.  For the
  Gauss--FFT method, $\FFTGs^{K}(m^2,r^2) = \FFTs^{K}(m^2,r^2) +
  2t\ceil{(t+1)/2}$, as we need two extra operations per complex
  number as described in Sec.~\ref{sub_sec_gauss}.

  All three methods need to fetch all kernel data from memory; read a
  total of $4CC'r^2$ bytes, and store $CC'$ transformed kernels back
  to main memory.  The transformed kernels have size $t \times t$, and
  are stored in the same fashion as the transformed input tiles,
  requiring total of $4t^2$, $8t\ceil{(t+1)/2}$, and
  $12t\ceil{(t+1)/2}$ bytes for the Winograd, Regular--FFT and
  Gauss--FFT methods respectively.

  The total number of operations, data movements, and \AI s for
  transforming all kernels of a particular layer, using any of the
  three methods, are given in Tbl.~\ref{table:layer-stuff}.

  \subsection{Element--wise Products} \label{sec_element_wise}
\begin{sloppypar}
  Having all transformed input and kernel tiles, the pre--transformed
  output tiles $\tilde{I'}_{b,c',n}$ are computed via \end{sloppypar}
  \begin{equation}
    \small %\scriptsize
    \begin{aligned}
      \tilde{I}'_{b,c',n} & = \sum_{c=1}^C \tilde{I}_{b,c,n} \odot \tilde{W}_{c,c'}
    \end{aligned}
    \label{eqn-ew-comp}
  \end{equation}
\begin{sloppypar}
  Here, all of the pre--transformed output tiles
  ($\tilde{I}'_{b,c',n}$), transformed input tiles
  $\tilde{I}_{b,c,n}$, and transformed kernels $\tilde{W}_{c,c'}$ have
  the same size $t \times t$.  Computing an element of the
  pre--transformed output tile at a location $\vec{e}$ depends only on
  elements of the transformed input tiles and transformed kernels at
  the same location, and is computed via:
\end{sloppypar}
  \begin{equation}
    \small %\scriptsize
    \begin{aligned}
      \tilde{I}'_{b,c',n}(\vec{e}) & = \sum_{c=1}^C \tilde{I}(\vec{e})_{b,c,n} \cdot \tilde{W}(\vec{e})_{c,c'}
    \end{aligned}
    \label{eqn-ew-comp}
  \end{equation}
  Note that the equation above can be interpreted as multiplication of
  a matrix $U_{\vec{e}}$ of size $BN \times C$ with a matrix
  $V_{\vec{e}}$ of size $C \times C'$ resulting in a matrix
  $X_{\vec{e}}$ of size $BN \times C'$.  For layers of modern
  ConvNets, values of $BN$ are much larger than $C$, which results in
  multiplication of tall and skinny
  matrices~\cite{winograd,lavin2016fast}. Such matrix multiplications
  have to be performed for each element location of
  $\tilde{I}'_{b,c',n}$.

  \subsubsection{Operations Required} \label{sub_sec_gemm_op}

  In the Winograd method, $t^2$ real matrices are multiplied,
  requiring a total of $2t^2BNCC'$ operations for the whole layer,
  where $N$ is the number of tiles per image.

  For the FFT based methods, complex matrices are multiplied.
  However, only $t \ceil{(t+1)/2}$ locations of the pre--transformed
  output tiles need to be computed.  As both the transformed input
  tiles and transformed kernels are conjugate symmetric, the
  pre--transformed output tiles will also be conjugate symmetric.

  In the Regular--FFT method, $4$ real multiply--adds are required for
  performing a complex multiply--add.  This gives us a total of $8t
  \ceil{(t+1)/2}BNCC'$ FLOPs required for the whole layer.  Gauss--FFT
  reduces complex matrix multiplications to 3 real matrix
  multiplications of matrices of the same size, totaling $6t
  \ceil{(t+1)/2}BNCC'$ FLOPs required for the whole layer.

  \subsubsection{Data Movement}

  In the Winograd method, $t^2$ real matrix multiplications $X = U
  \times V$ are performed (for the rest of the section we will omit
  the subscript $_{\vec{e}}$, for clarity).  In the Regular--FFT
  method, $t\ceil{(t+1)/2}$ complex matrix multiplications are
  performed, with the same sizes.  The Gauss--FFT method replaces one
  complex matrix multiplication with three real matrix
  multiplications; thus, a total of $3t\ceil{(t+1)/2}$ real matrix
  multiplications are performed.

  For modern ConvNets, the matrices $X$, $U$ and $V$ may not fit in
  cache, in which case standard cache--blocking approaches are
  employed~\cite{Gannon:1987:SCL:647970.761024, heinecke2015libxsmm,
    heinecke2016libxsmm, winograd, li2015input}, and might require
  some data to be moved to and/or from the main memory multiple times.

  In the standard cache--blocking technique, $V$ (of size $C \times
  C'$) is subdivided into equally sized matrices of size $c
  \times c'$, 

  To minimize transfers to and from main memory, a sub--matrix of $V$ 
  is kept in cache. A small sub--matrix of $U$, with size of $\rho \stimes c$, is fetched from memory,
  multiplied with the sub--matrix of $V$ stored in--cache, and accumulated
  to the appropriate sub--matrix of $X$.  Here $\rho$ is a small number, required
  for efficient in--cache computation~\cite{Gannon:1987:SCL:647970.761024, heinecke2015libxsmm,
    heinecke2016libxsmm, winograd, li2015input}.
  
  This requires transferring a total of $\rho c$
  numbers from main memory, and $\rho c'$ numbers bytes of $X$ from
  and then back to the main memory.  A total of $\rho(c + 2c')$
  numbers.  In the special case, when $c = C$, only $\rho(c + c')$
  numbers need to be moved, as each sub--matrix multiplication 
  produces the final result (no accumulation is necessary).
  
  Total of $\nicefrac{BNCC'}{\rho cc'}$ such sub--matrix multiplications
  are performed, and require $\nicefrac{BNCC'}{cc'}(c + \alpha c')$ numbers
  to be moved. With $\alpha$ being $1$ when $c = C$ and $2$ when $c < C$.

  In the Winograd method $t^2$ real matrix multiplications ($X = U
  \times V$) are performed, thus transferring a total of $4t^2BN(c +
  \alpha c')\nicefrac{CC'}{cc'}$ bytes to be moved.

  In the Regular--FFT method $t\ceil{(t+1)/2}$ complex matrix
  multiplications are performed, requiring a total of
  $8t\ceil{(t+1)/2}BN(c + \alpha c')\nicefrac{CC'}{cc'}$ bytes to be
  moved.

  The Gauss--FFT replaces each of the $t\ceil{(t+1)/2}$ complex matrix
  multiplications with 3 real matrix multiplications.  Total of
  $12t\ceil{(t+1)/2}BN(c + \alpha c')\nicefrac{CC'}{cc'}$ bytes need
  to be transferred.

  The total number of operations is fixed.  To minimize the amount of data
  movements we need to choose optimal values for $c$ and $c'$, while allowing for in--cache computation

  As the values of $t$, $B$, $C$, $C'$ and $N$ are constant, the
  optimal values of $c$ and $c'$ can be chosen by solving the
  following optimization problem:
  \begin{equation} \label{eq:gemm_AI}
    \small
    \begin{aligned}
      & \underset{\mathbf{c,c'}}{\text{minimize}} &&
      \frac{(c + \alpha c')}{cc'} \\
      & \text{subject to}
      & & \mathbf{c} \mid C & \text{($C$ is divisible by $\mathbf{c}$)}\\
      & & & \mathbf{c'} \mid C'& \text{($C'$ is divisible by $\mathbf{c'}$)} \\
      & & & \mathbf{4\beta cc'} \le \frac{\text{Cache Size}}{2} & \text{(fits in half cache)}
    \end{aligned}
  \end{equation}
  Where $\alpha$ equals $1$ when $c = C$ and $2$ when $c < C$; $\beta$
  is $1$ for real valued matrices, and $2$ for complex valued ones.
  Half the cache is allowed for sub--matrices of $V$.  This is typical
  practice, to ensure enough space for sub--matrices of $U$ and $X$.

  \begin{figure}
    \begin{center}
      \includegraphics[width=0.88\linewidth]{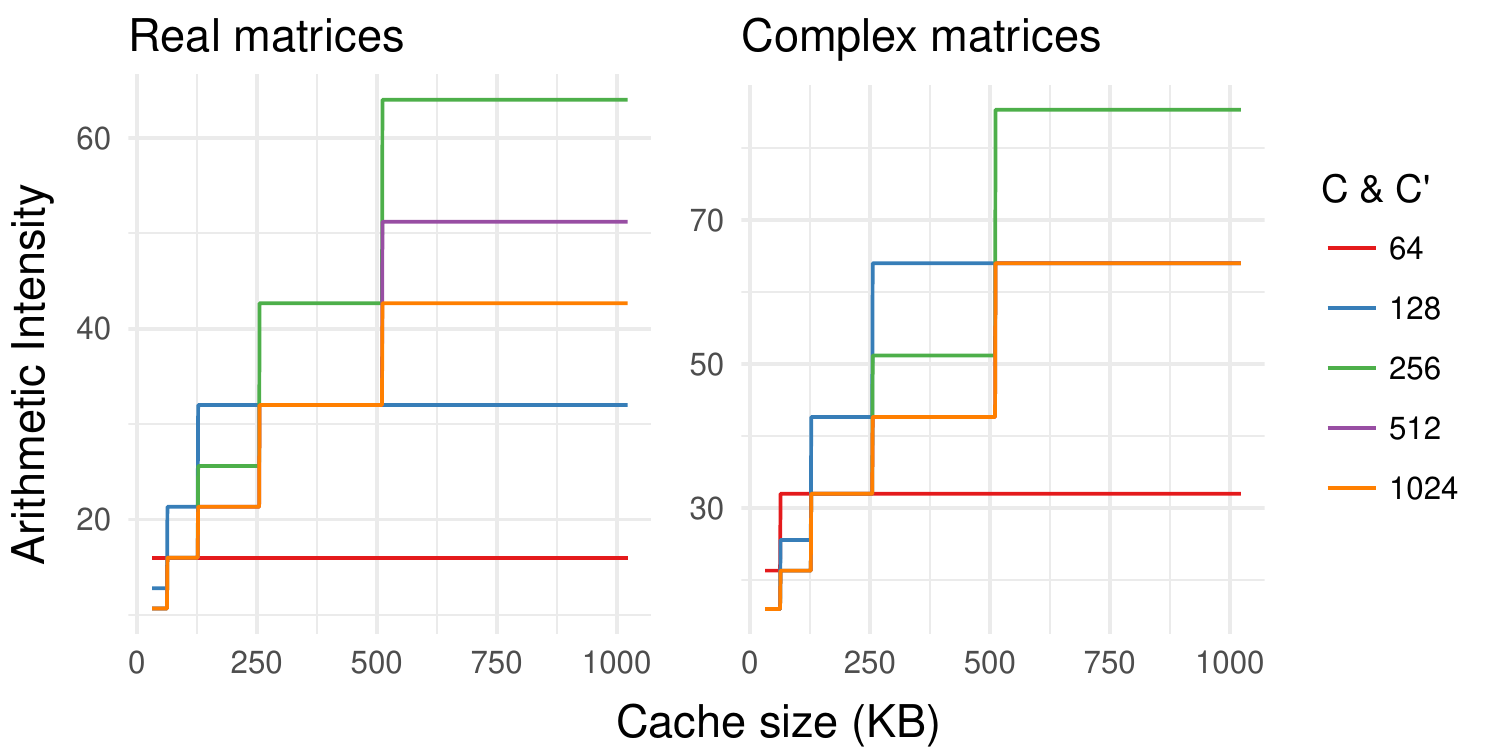}
    \end{center}
    \caption{Arithmetic intensities for the element--wise stage given
      various amounts of cache.  In the Winograd and Gauss--FFT
      method, real matrix multiplications are performed, whereas, in
      the Regular--FFT complex matrix multiplications are performed.}
    \label{fig:gemm-cgemm-ratios}
  \end{figure}

  The arithmetic intensities can be then computed by dividing the
  number of required operations with the amount of required data
  movements for optimal values of $c$ and $c'$.  This results in the \AI s of
  $cc' / 2(c + \alpha c')$ for the Winograd and Gauss--FFT methods,
  and $cc' / (c + \alpha c')$ for the Regular--FFT method, which are equal
  to the \AI s of real matrix, and complex matrix multiplications,
  respectively~\cite{Gannon:1987:SCL:647970.761024,
    heinecke2015libxsmm, heinecke2016libxsmm, winograd}.

  In Fig.~\ref{fig:gemm-cgemm-ratios}, we show the arithmetic
  intensities for layers with different numbers of channels as a
  function of cache size.  The \AI s of both complex (used in
  Regular--FFT method) and real matrix multiplication (used in
  Winograd and Gauss--FFT) increase with the cache size and the number
  of channels (C and C').  For a fixed cache size, the complex matrix
  multiplication allows for a higher arithmetic intensity.  This
  indicates that the element--wise stage of the Regular--FFT method
  may achieve better hardware utilization than the ones of Winograd or
  Gauss--FFT convolution, when the cache size is a limiting resource.

  \subsection{Output Transform} \label{subsec_inverse_T}
  
\begin{sloppypar}
  In the output transform stage, each of the $BNC'$ pre--transformed
  output tiles $\tilde{I'}_{b,c',n}$ is transformed from the Winograd/FFT
  domain back to the spatial domain via $I'_{b,c,n} =
  \mathbb{A}^T\tilde{I}'_{b,c,n}\mathbb{A}$.  The matrix $\mathbb{A}$
  has size of $t \times m$ ($t = m + r -1$), resulting in
  $I'_{b,c',n}$ having a size of $m \times m$.  Here $I'_{b,c',n}$ is
  the $n$--th (non--overlapping) tile of the $c'$-th image in batch
  $b$ of the final result of the convolutional layer.
\end{sloppypar}

  Performing the inverse transforms is very similar to performing the input
  transforms, except that:  1) Only a subset of $m \stimes m$ elements
  need to be computed, and 2) Different, (inverse transform) matrices 
  (or inverse FFT transforms) are used.

  Given the analysis of the input and kernel transforms, analyzing
  the output transforms is straight--forward.  The total number of 
  operations, data movements, and \AI s for
  transforming output tiles of a particular layer, using any of the
  three methods are given in Tbl.~\ref{table:layer-stuff}.

%\appendix

 \section{Model Lookup Tables}

    \begin{table*} \centering
    \small
    \setlength\tabcolsep{2.5pt}
    \caption{FLOPS required for performing Winograd transforms of a single tile or kernel.}
    \begin{tabular}{c | ccc | ccc | ccc | ccc | ccc | ccc}
      \toprule
      & \multicolumn{3}{c|}{$r = 2$}& \multicolumn{3}{c|}{$r = 3$}& \multicolumn{3}{c|}{$r = 4$}& \multicolumn{3}{c|}{$r = 5$}& \multicolumn{3}{c|}{$r = 6$}& \multicolumn{3}{c}{$r = 7$} \\
      Method & In & Ker & Out  & In & Ker & Out  & In & Ker & Out  & In & Ker & Out  & In & Ker & Out  & In & Ker & Out \\
      \midrule
      $\WINs(2^2,r^2)$  & 12 & 5 & 5 & 32 & 49 & 42 & 180 & 198 & 154 & 240 & 231 & 168 & 742 & 728 & 504 & 704 & 645 & 430 \\
      $\WINs(3^2,r^2)$  & 32 & 24 & 28 & 180 & 128 & 128 & 240 & 170 & 153 & 742 & 552 & 460 & 704 & 518 & 407 & - & - & - \\
      $\WINs(4^2,r^2)$  & 180 & 70 & 90 & 240 & 135 & 150 & 742 & 396 & 396 & 704 & 403 & 372 & - & - & - & - & - & - \\
      $\WINs(5^2,r^2)$  & 240 & 72 & 99 & 742 & 260 & 312 & 704 & 300 & 325 & - & - & - & - & - & - & - & - & - \\
      $\WINs(6^2,r^2)$  & 742 & 144 & 208 & 704 & 242 & 308 & - & - & - & - & - & - & - & - & - & - & - & - \\
      $\WINs(7^2,r^2)$  & 704 & 130 & 195 & - & - & - & - & - & - & - & - & - & - & - & - & - & - & - \\
      \bottomrule
    \end{tabular}
    \label{table:winflops}
  \end{table*}

  \begin{table*} \centering
    \small
    \setlength\tabcolsep{2.5pt}
    \caption{AIs of Winograd transforms for a single tile or kernel.}
    \begin{tabular}{c | ccc | ccc | ccc | ccc | ccc | ccc}
      \toprule
      & \multicolumn{3}{c|}{$r = 2$}& \multicolumn{3}{c|}{$r = 3$}& \multicolumn{3}{c|}{$r = 4$}& \multicolumn{3}{c|}{$r = 5$}& \multicolumn{3}{c|}{$r = 6$}& \multicolumn{3}{c}{$r = 7$} \\
      Method & In & Ker & Out  & In & Ker & Out  & In & Ker & Out  & In & Ker & Out  & In & Ker & Out  & In & Ker & Out \\
      \midrule

$\WINs(2^2,r^2)$  & 0.17 & 0.10 & 0.10 & 0.25 & 0.49 & 0.53 & 0.90 & 1.21 & 1.33 & 0.83 & 0.95 & 1.05 & 1.89 & 2.14 & 2.38 & 1.38 & 1.43 & 1.58 \\
$\WINs(3^2,r^2)$  & 0.25 & 0.30 & 0.28 & 0.90 & 0.94 & 0.94 & 0.83 & 0.82 & 0.85 & 1.89 & 1.86 & 1.98 & 1.38 & 1.29 & 1.39 & - & - & - \\
$\WINs(4^2,r^2)$  & 0.90 & 0.60 & 0.55 & 0.83 & 0.75 & 0.72 & 1.89 & 1.52 & 1.52 & 1.38 & 1.13 & 1.16 & - & - & - & - & - & - \\
$\WINs(5^2,r^2)$  & 0.83 & 0.45 & 0.41 & 1.89 & 1.12 & 1.05 & 1.38 & 0.94 & 0.91 & - & - & - & - & - & - & - & - & - \\
$\WINs(6^2,r^2)$  & 1.89 & 0.68 & 0.61 & 1.38 & 0.83 & 0.77 & - & - & - & - & - & - & - & - & - & - & - & - \\
$\WINs(7^2,r^2)$  & 1.38 & 0.48 & 0.43 & - & - & - & - & - & - & - & - & - & - & - & - & - & - & - \\
      \bottomrule
    \end{tabular}
    \label{table:winai}
  \end{table*}

  \begin{table*} \centering
    \small
    \setlength\tabcolsep{2.5pt}
    \caption{FLOPS required for performing Regular--FFT transforms of a single tile or kernel.}
    \begin{tabular}{c | ccc | ccc | ccc | ccc | ccc | ccc}
      \toprule
      & \multicolumn{3}{c|}{$r = 2$}& \multicolumn{3}{c|}{$r = 3$}& \multicolumn{3}{c|}{$r = 4$}& \multicolumn{3}{c|}{$r = 5$}& \multicolumn{3}{c|}{$r = 6$}& \multicolumn{3}{c}{$r = 7$} \\
      Method & In & Ker & Out  & In & Ker & Out  & In & Ker & Out  & In & Ker & Out  & In & Ker & Out  & In & Ker & Out \\
      \midrule
      $\FFTs(2^2,r^2)$  & 54 & 36 & 42 & 72 & 48 & 48 & 245 & 206 & 152 & 300 & 256 & 160 & 702 & 632 & 288 & 492 & 434 & 191 \\
      $\FFTs(3^2,r^2)$  & 72 & 28 & 66 & 245 & 171 & 186 & 300 & 192 & 197 & 702 & 566 & 444 & 492 & 380 & 263 & 1292 & 1088 & 678 \\
      $\FFTs(4^2,r^2)$  & 245 & 92 & 224 & 300 & 158 & 232 & 702 & 504 & 504 & 492 & 320 & 335 & 1292 & 992 & 758 & 1200 & 904 & 768 \\
      $\FFTs(5^2,r^2)$  & 300 & 128 & 274 & 702 & 325 & 568 & 492 & 276 & 400 & 1292 & 805 & 850 & 1200 & 792 & 832 & 2710 & 2118 & 1820 \\
      $\FFTs(6^2,r^2)$  & 702 & 172 & 636 & 492 & 206 & 453 & 1292 & 624 & 930 & 1200 & 722 & 912 & 2710 & 1980 & 1956 & 1416 & 924 & 1074 \\
      $\FFTs(7^2,r^2)$  & 492 & 146 & 503 & 1292 & 491 & 1014 & 1200 & 656 & 996 & 2710 & 1523 & 2096 & 1416 & 896 & 1141 & 3681 & 2583 & 2646 \\
      $\FFTs(8^2,r^2)$  & 1292 & 296 & 1102 & 1200 & 564 & 1084 & 2710 & 1108 & 2240 & 1416 & 823 & 1190 & 3681 & 2432 & 2808 & 3228 & 2082 & 2384 \\
      $\FFTs(9^2,r^2)$  & 1200 & 340 & 1176 & 2710 & 735 & 2388 & 1416 & 672 & 1262 & 3681 & 2304 & 3066 & 3228 & 1960 & 2526 & 3405 & 2254 & 2499 \\
      $\FFTs(10^2,r^2)$  & 2710 & 404 & 2540 & 1416 & 610 & 1336 & 3681 & 2052 & 3240 & 3228 & 1842 & 2672 & 3405 & 2082 & 2616 & 2904 & 1688 & 2292 \\
      $\FFTs(11^2,r^2)$  & 1416 & 364 & 1412 & 3681 & 1663 & 3418 & 3228 & 1504 & 2822 & 3405 & 1865 & 2897 & 2904 & 1530 & 2428 & 7793 & 5153 & 6354 \\
      $\FFTs(12^2,r^2)$  & 3681 & 724 & 3600 & 3228 & 1162 & 2976 & 3405 & 1628 & 3052 & 2904 & 1374 & 2562 & 7793 & 4716 & 6630 & 6068 & 3479 & 4228 \\
      $\FFTs(13^2,r^2)$  & 3228 & 648 & 3134 & 3405 & 1324 & 3211 & 2904 & 1246 & 2707 & 7793 & 4341 & 6892 & 6068 & 3290 & 4434 & 14418 & 8629 & 11023 \\
      $\FFTs(14^2,r^2)$  & 3405 & 764 & 3374 & 2904 & 936 & 2824 & 7793 & 4058 & 7167 & 6068 & 2660 & 4648 & 14418 & 8176 & 11410 & 5440 & 2876 & 4512 \\
      $\FFTs(15^2,r^2)$  & 2904 & 634 & 2976 & 7793 & 3231 & 7446 & 6068 & 1944 & 4840 & 14418 & 7305 & 11775 & 5440 & 2580 & 4682 & 8658 & 5617 & 6841 \\
      $\FFTs(16^2,r^2)$  & 7793 & 1616 & 7720 & 6068 & 1540 & 5036 & 14418 & 6668 & 12288 & 5440 & 2409 & 4878 & 8658 & 4880 & 7470 & 11852 & 5742 & 9976 \\
      $\FFTs(17^2,r^2)$  & 6068 & 1132 & 5236 & 14418 & 6266 & 12699 & 5440 & 2204 & 5078 & 8658 & 4515 & 7733 & 11852 & 4920 & 10296 & 26954 & 15562 & 23206 \\
      $\FFTs(18^2,r^2)$  & 14418 & 4516 & 13082 & 5440 & 1919 & 5282 & 8658 & 3630 & 8000 & 11852 & 4268 & 10620 & 26954 & 15072 & 23808 & 7524 & 3580 & 6622 \\
      $\FFTs(19^2,r^2)$  & 5440 & 1196 & 5490 & 8658 & 2557 & 8271 & 11852 & 3432 & 10948 & 26954 & 11815 & 24414 & 7524 & 3440 & 6837 & 17260 & 7145 & 12714 \\
      $\FFTs(20^2,r^2)$  & 8658 & 1448 & 8546 & 11852 & 2718 & 11280 & 26954 & 10880 & 25024 & 7524 & 3104 & 6988 & 17260 & 6432 & 13066 & 16128 & 8218 & 14180 \\
      $\FFTs(21^2,r^2)$  & 11852 & 1552 & 11616 & 26954 & 9981 & 25638 & 7524 & 2612 & 7213 & 17260 & 5775 & 13422 & 16128 & 7944 & 14560 & 21050 & 8309 & 15183 \\
      $\FFTs(22^2,r^2)$  & 26954 & 8396 & 26624 & 7524 & 2176 & 7363 & 17260 & 4746 & 13782 & 16128 & 7146 & 14944 & 21050 & 7586 & 15566 & 14184 & 6929 & 12426 \\
      $\FFTs(23^2,r^2)$  & 7524 & 1442 & 7570 & 17260 & 3773 & 14146 & 16128 & 6544 & 15332 & 21050 & 6308 & 15953 & 14184 & 6552 & 12760 & 34745 & 18690 & 31062 \\
      $\FFTs(24^2,r^2)$  & 17260 & 2076 & 14514 & 16128 & 6216 & 15724 & 21050 & 5100 & 16344 & 14184 & 6205 & 13098 & 34745 & 17994 & 31704 & 15480 & 7333 & 13912 \\
      $\FFTs(25^2,r^2)$  & 16128 & 2596 & 16120 & 21050 & 4118 & 16739 & 14184 & 5624 & 13440 & 34745 & 17385 & 32350 & 15480 & 6704 & 14284 & 38400 & 19470 & 34934 \\
      $\FFTs(26^2,r^2)$  & 21050 & 2516 & 17138 & 14184 & 4089 & 13786 & 34745 & 15508 & 33000 & 15480 & 6218 & 14610 & 38400 & 19142 & 35600 & 15804 & 7224 & 14352 \\
      $\FFTs(27^2,r^2)$  & 14184 & 2356 & 14136 & 34745 & 14655 & 33654 & 15480 & 4856 & 14933 & 38400 & 18275 & 36270 & 15804 & 6574 & 14669 & - & - & - \\
      $\FFTs(28^2,r^2)$  & 34745 & 11276 & 34544 & 15480 & 3842 & 15292 & 38400 & 17596 & 36944 & 15804 & 5924 & 14962 & - & - & - & - & - & - \\
      $\FFTs(29^2,r^2)$  & 15480 & 2840 & 15655 & 38400 & 14293 & 37622 & 15804 & 5302 & 15286 & - & - & - & - & - & - & - & - & - \\
      $\FFTs(30^2,r^2)$  & 38400 & 10940 & 38394 & 15804 & 3856 & 15648 & - & - & - & - & - & - & - & - & - & - & - & - \\
      $\FFTs(31^2,r^2)$  & 15804 & 2570 & 16014 & - & - & - & - & - & - & - & - & - & - & - & - & - & - & - \\
      \bottomrule
    \end{tabular}
    \label{table:fftflops}
  \end{table*}

    \begin{table*} \centering
    \small
    \setlength\tabcolsep{2.5pt}
    \caption{AIs of Regular--FFT transforms for a single tile or kernel.}
    \begin{tabular}{c | ccc | ccc | ccc | ccc | ccc | ccc}
      \toprule
      & \multicolumn{3}{c|}{$r = 2$}& \multicolumn{3}{c|}{$r = 3$}& \multicolumn{3}{c|}{$r = 4$}& \multicolumn{3}{c|}{$r = 5$}& \multicolumn{3}{c|}{$r = 6$}& \multicolumn{3}{c}{$r = 7$} \\
      Method & In & Ker & Out  & In & Ker & Out  & In & Ker & Out  & In & Ker & Out  & In & Ker & Out  & In & Ker & Out \\
      \midrule

$\FFTs(2^2,r^2)$  & 0.64 & 0.56 & 0.66 & 0.45 & 0.36 & 0.43 & 1.11 & 1.12 & 1.12 & 0.89 & 0.88 & 0.77 & 1.67 & 1.72 & 1.20 & 0.85 & 0.84 & 0.57 \\
$\FFTs(3^2,r^2)$  & 0.45 & 0.25 & 0.50 & 1.11 & 1.10 & 1.19 & 0.89 & 0.75 & 0.86 & 1.67 & 1.75 & 1.71 & 0.85 & 0.82 & 0.74 & 1.89 & 1.96 & 1.71 \\
$\FFTs(4^2,r^2)$  & 1.11 & 0.68 & 1.22 & 0.89 & 0.69 & 0.91 & 1.67 & 1.75 & 1.75 & 0.85 & 0.76 & 0.87 & 1.89 & 1.97 & 1.79 & 1.36 & 1.34 & 1.41 \\
$\FFTs(5^2,r^2)$  & 0.89 & 0.62 & 0.94 & 1.67 & 1.25 & 1.75 & 0.85 & 0.72 & 0.95 & 1.89 & 1.75 & 1.85 & 1.36 & 1.27 & 1.43 & 2.68 & 2.93 & 2.90 \\
$\FFTs(6^2,r^2)$  & 1.67 & 0.72 & 1.73 & 0.85 & 0.58 & 0.98 & 1.89 & 1.47 & 1.85 & 1.36 & 1.24 & 1.46 & 2.68 & 2.95 & 2.91 & 1.13 & 1.06 & 1.32 \\
$\FFTs(7^2,r^2)$  & 0.85 & 0.43 & 0.97 & 1.89 & 1.24 & 1.82 & 1.36 & 1.21 & 1.47 & 2.68 & 2.43 & 2.90 & 1.13 & 1.10 & 1.31 & 2.62 & 2.80 & 2.86 \\
$\FFTs(8^2,r^2)$  & 1.89 & 0.79 & 1.79 & 1.36 & 1.09 & 1.47 & 2.68 & 1.87 & 2.86 & 1.13 & 1.07 & 1.28 & 2.62 & 2.79 & 2.85 & 1.92 & 1.91 & 2.07 \\
$\FFTs(9^2,r^2)$  & 1.36 & 0.69 & 1.46 & 2.68 & 1.30 & 2.80 & 1.13 & 0.91 & 1.27 & 2.62 & 2.78 & 2.91 & 1.92 & 1.88 & 2.07 & 1.83 & 1.95 & 1.95 \\
$\FFTs(10^2,r^2)$  & 2.68 & 0.74 & 2.74 & 1.13 & 0.86 & 1.25 & 2.62 & 2.59 & 2.87 & 1.92 & 1.85 & 2.06 & 1.83 & 1.89 & 1.92 & 1.33 & 1.25 & 1.48 \\
$\FFTs(11^2,r^2)$  & 1.13 & 0.53 & 1.22 & 2.62 & 2.18 & 2.82 & 1.92 & 1.57 & 2.04 & 1.83 & 1.76 & 2.01 & 1.33 & 1.18 & 1.48 & 3.27 & 3.63 & 3.72 \\
$\FFTs(12^2,r^2)$  & 2.62 & 0.97 & 2.76 & 1.92 & 1.25 & 2.02 & 1.83 & 1.59 & 1.99 & 1.33 & 1.10 & 1.48 & 3.27 & 3.45 & 3.68 & 2.22 & 2.13 & 2.10 \\
$\FFTs(13^2,r^2)$  & 1.92 & 0.71 & 1.99 & 1.83 & 1.33 & 1.96 & 1.33 & 1.02 & 1.48 & 3.27 & 3.28 & 3.63 & 2.22 & 2.08 & 2.10 & 4.86 & 5.03 & 5.02 \\
$\FFTs(14^2,r^2)$  & 1.83 & 0.78 & 1.93 & 1.33 & 0.79 & 1.46 & 3.27 & 3.15 & 3.57 & 2.22 & 1.73 & 2.09 & 4.86 & 4.91 & 4.95 & 1.62 & 1.47 & 1.77 \\
$\FFTs(15^2,r^2)$  & 1.33 & 0.54 & 1.45 & 3.27 & 2.56 & 3.51 & 2.22 & 1.29 & 2.07 & 4.86 & 4.51 & 4.87 & 1.62 & 1.36 & 1.76 & 2.40 & 2.75 & 2.49 \\
$\FFTs(16^2,r^2)$  & 3.27 & 1.30 & 3.43 & 2.22 & 1.04 & 2.04 & 4.86 & 4.21 & 4.83 & 1.62 & 1.30 & 1.75 & 2.40 & 2.45 & 2.60 & 2.93 & 2.49 & 3.18 \\
$\FFTs(17^2,r^2)$  & 2.22 & 0.78 & 2.02 & 4.86 & 4.03 & 4.75 & 1.62 & 1.21 & 1.74 & 2.40 & 2.32 & 2.57 & 2.93 & 2.18 & 3.15 & 6.23 & 6.47 & 6.90 \\
$\FFTs(18^2,r^2)$  & 4.86 & 2.94 & 4.65 & 1.62 & 1.07 & 1.73 & 2.40 & 1.90 & 2.54 & 2.93 & 1.93 & 3.12 & 6.23 & 6.41 & 6.79 & 1.57 & 1.33 & 1.75 \\
$\FFTs(19^2,r^2)$  & 1.62 & 0.67 & 1.71 & 2.40 & 1.36 & 2.51 & 2.93 & 1.58 & 3.08 & 6.23 & 5.12 & 6.69 & 1.57 & 1.30 & 1.74 & 3.38 & 2.56 & 3.14 \\
$\FFTs(20^2,r^2)$  & 2.40 & 0.78 & 2.48 & 2.93 & 1.27 & 3.04 & 6.23 & 4.79 & 6.57 & 1.57 & 1.20 & 1.71 & 3.38 & 2.34 & 3.11 & 2.87 & 2.64 & 3.14 \\
$\FFTs(21^2,r^2)$  & 2.93 & 0.73 & 3.00 & 6.23 & 4.45 & 6.45 & 1.57 & 1.02 & 1.69 & 3.38 & 2.14 & 3.08 & 2.87 & 2.60 & 3.11 & 3.54 & 2.58 & 3.17 \\
$\FFTs(22^2,r^2)$  & 6.23 & 3.78 & 6.42 & 1.57 & 0.86 & 1.66 & 3.38 & 1.78 & 3.04 & 2.87 & 2.37 & 3.08 & 3.54 & 2.39 & 3.14 & 2.18 & 1.95 & 2.35 \\
$\FFTs(23^2,r^2)$  & 1.57 & 0.57 & 1.64 & 3.38 & 1.43 & 3.00 & 2.87 & 2.20 & 3.05 & 3.54 & 2.02 & 3.10 & 2.18 & 1.87 & 2.33 & 5.08 & 5.08 & 5.55 \\
$\FFTs(24^2,r^2)$  & 3.38 & 0.79 & 2.96 & 2.87 & 2.11 & 3.01 & 3.54 & 1.65 & 3.07 & 2.18 & 1.79 & 2.31 & 5.08 & 4.97 & 5.48 & 2.08 & 1.82 & 2.26 \\
$\FFTs(25^2,r^2)$  & 2.87 & 0.89 & 2.98 & 3.54 & 1.35 & 3.03 & 2.18 & 1.64 & 2.29 & 5.08 & 4.86 & 5.41 & 2.08 & 1.68 & 2.25 & 4.92 & 4.68 & 5.40 \\
$\FFTs(26^2,r^2)$  & 3.54 & 0.83 & 2.99 & 2.18 & 1.20 & 2.27 & 5.08 & 4.38 & 5.34 & 2.08 & 1.58 & 2.23 & 4.92 & 4.66 & 5.34 & 1.87 & 1.59 & 2.03 \\
$\FFTs(27^2,r^2)$  & 2.18 & 0.70 & 2.25 & 5.08 & 4.17 & 5.26 & 2.08 & 1.24 & 2.21 & 4.92 & 4.49 & 5.27 & 1.87 & 1.46 & 2.02 & - & - & - \\
$\FFTs(28^2,r^2)$  & 5.08 & 3.23 & 5.22 & 2.08 & 0.99 & 2.19 & 4.92 & 4.36 & 5.20 & 1.87 & 1.33 & 2.00 & - & - & - & - & - & - \\
$\FFTs(29^2,r^2)$  & 2.08 & 0.74 & 2.17 & 4.92 & 3.57 & 5.13 & 1.87 & 1.20 & 1.98 & - & - & - & - & - & - & - & - & - \\
$\FFTs(30^2,r^2)$  & 4.92 & 2.75 & 5.07 & 1.87 & 0.88 & 1.97 & - & - & - & - & - & - & - & - & - & - & - & - \\
$\FFTs(31^2,r^2)$  & 1.87 & 0.59 & 1.95 & - & - & - & - & - & - & - & - & - & - & - & - & - & - & - \\
      \bottomrule
    \end{tabular}
    \label{table:fftai}
  \end{table*}

  \begin{table*} \centering
    \small
    \setlength\tabcolsep{2.5pt}
    \caption{FLOPS required for performing Gauss--FFT transforms of a single tile or kernel.}
    \begin{tabular}{c | ccc | ccc | ccc | ccc | ccc | ccc}
      \toprule
      & \multicolumn{3}{c|}{$r = 2$}& \multicolumn{3}{c|}{$r = 3$}& \multicolumn{3}{c|}{$r = 4$}& \multicolumn{3}{c|}{$r = 5$}& \multicolumn{3}{c|}{$r = 6$}& \multicolumn{3}{c}{$r = 7$} \\
      Method & In & Ker & Out  & In & Ker & Out  & In & Ker & Out  & In & Ker & Out  & In & Ker & Out  & In & Ker & Out \\
      \midrule

$\FFTGs(2^2,r^2)$  & 63 & 54 & 60 & 88 & 80 & 80 & 270 & 256 & 202 & 336 & 328 & 232 & 751 & 730 & 386 & 556 & 562 & 319 \\
$\FFTGs(3^2,r^2)$  & 88 & 60 & 98 & 270 & 221 & 236 & 336 & 264 & 269 & 751 & 664 & 542 & 556 & 508 & 391 & 1373 & 1250 & 840 \\
$\FFTGs(4^2,r^2)$  & 270 & 142 & 274 & 336 & 230 & 304 & 751 & 602 & 602 & 556 & 448 & 463 & 1373 & 1154 & 920 & 1300 & 1104 & 968 \\
$\FFTGs(5^2,r^2)$  & 336 & 200 & 346 & 751 & 423 & 666 & 556 & 404 & 528 & 1373 & 967 & 1012 & 1300 & 992 & 1032 & 2831 & 2360 & 2062 \\
$\FFTGs(6^2,r^2)$  & 751 & 270 & 734 & 556 & 334 & 581 & 1373 & 786 & 1092 & 1300 & 922 & 1112 & 2831 & 2222 & 2198 & 1560 & 1212 & 1362 \\
$\FFTGs(7^2,r^2)$  & 556 & 274 & 631 & 1373 & 653 & 1176 & 1300 & 856 & 1196 & 2831 & 1765 & 2338 & 1560 & 1184 & 1429 & 3850 & 2921 & 2984 \\
$\FFTGs(8^2,r^2)$  & 1373 & 458 & 1264 & 1300 & 764 & 1284 & 2831 & 1350 & 2482 & 1560 & 1111 & 1478 & 3850 & 2770 & 3146 & 3424 & 2474 & 2776 \\
$\FFTGs(9^2,r^2)$  & 1300 & 540 & 1376 & 2831 & 977 & 2630 & 1560 & 960 & 1550 & 3850 & 2642 & 3404 & 3424 & 2352 & 2918 & 3630 & 2704 & 2949 \\
$\FFTGs(10^2,r^2)$  & 2831 & 646 & 2782 & 1560 & 898 & 1624 & 3850 & 2390 & 3578 & 3424 & 2234 & 3064 & 3630 & 2532 & 3066 & 3160 & 2200 & 2804 \\
$\FFTGs(11^2,r^2)$  & 1560 & 652 & 1700 & 3850 & 2001 & 3756 & 3424 & 1896 & 3214 & 3630 & 2315 & 3347 & 3160 & 2042 & 2940 & 8082 & 5731 & 6932 \\
$\FFTGs(12^2,r^2)$  & 3850 & 1062 & 3938 & 3424 & 1554 & 3368 & 3630 & 2078 & 3502 & 3160 & 1886 & 3074 & 8082 & 5294 & 7208 & 6392 & 4127 & 4876 \\
$\FFTGs(13^2,r^2)$  & 3424 & 1040 & 3526 & 3630 & 1774 & 3661 & 3160 & 1758 & 3219 & 8082 & 4919 & 7470 & 6392 & 3938 & 5082 & 14779 & 9351 & 11745 \\
$\FFTGs(14^2,r^2)$  & 3630 & 1214 & 3824 & 3160 & 1448 & 3336 & 8082 & 4636 & 7745 & 6392 & 3308 & 5296 & 14779 & 8898 & 12132 & 5840 & 3676 & 5312 \\
$\FFTGs(15^2,r^2)$  & 3160 & 1146 & 3488 & 8082 & 3809 & 8024 & 6392 & 2592 & 5488 & 14779 & 8027 & 12497 & 5840 & 3380 & 5482 & 9099 & 6499 & 7723 \\
$\FFTGs(16^2,r^2)$  & 8082 & 2194 & 8298 & 6392 & 2188 & 5684 & 14779 & 7390 & 13010 & 5840 & 3209 & 5678 & 9099 & 5762 & 8352 & 12336 & 6710 & 10944 \\
$\FFTGs(17^2,r^2)$  & 6392 & 1780 & 5884 & 14779 & 6988 & 13421 & 5840 & 3004 & 5878 & 9099 & 5397 & 8615 & 12336 & 5888 & 11264 & 27483 & 16620 & 24264 \\
$\FFTGs(18^2,r^2)$  & 14779 & 5238 & 13804 & 5840 & 2719 & 6082 & 9099 & 4512 & 8882 & 12336 & 5236 & 11588 & 27483 & 16130 & 24866 & 8100 & 4732 & 7774 \\
$\FFTGs(19^2,r^2)$  & 5840 & 1996 & 6290 & 9099 & 3439 & 9153 & 12336 & 4400 & 11916 & 27483 & 12873 & 25472 & 8100 & 4592 & 7989 & 17885 & 8395 & 13964 \\
$\FFTGs(20^2,r^2)$  & 9099 & 2330 & 9428 & 12336 & 3686 & 12248 & 27483 & 11938 & 26082 & 8100 & 4256 & 8140 & 17885 & 7682 & 14316 & 16804 & 9570 & 15532 \\
$\FFTGs(21^2,r^2)$  & 12336 & 2520 & 12584 & 27483 & 11039 & 26696 & 8100 & 3764 & 8365 & 17885 & 7025 & 14672 & 16804 & 9296 & 15912 & 21779 & 9767 & 16641 \\
$\FFTGs(22^2,r^2)$  & 27483 & 9454 & 27682 & 8100 & 3328 & 8515 & 17885 & 5996 & 15032 & 16804 & 8498 & 16296 & 21779 & 9044 & 17024 & 14968 & 8497 & 13994 \\
$\FFTGs(23^2,r^2)$  & 8100 & 2594 & 8722 & 17885 & 5023 & 15396 & 16804 & 7896 & 16684 & 21779 & 7766 & 17411 & 14968 & 8120 & 14328 & 35586 & 20372 & 32744 \\
$\FFTGs(24^2,r^2)$  & 17885 & 3326 & 15764 & 16804 & 7568 & 17076 & 21779 & 6558 & 17802 & 14968 & 7773 & 14666 & 35586 & 19676 & 33386 & 16380 & 9133 & 15712 \\
$\FFTGs(25^2,r^2)$  & 16804 & 3948 & 17472 & 21779 & 5576 & 18197 & 14968 & 7192 & 15008 & 35586 & 19067 & 34032 & 16380 & 8504 & 16084 & 39361 & 21392 & 36856 \\
$\FFTGs(26^2,r^2)$  & 21779 & 3974 & 18596 & 14968 & 5657 & 15354 & 35586 & 17190 & 34682 & 16380 & 8018 & 16410 & 39361 & 21064 & 37522 & 16828 & 9272 & 16400 \\
$\FFTGs(27^2,r^2)$  & 14968 & 3924 & 15704 & 35586 & 16337 & 35336 & 16380 & 6656 & 16733 & 39361 & 20197 & 38192 & 16828 & 8622 & 16717 & - & - & - \\
$\FFTGs(28^2,r^2)$  & 35586 & 12958 & 36226 & 16380 & 5642 & 17092 & 39361 & 19518 & 38866 & 16828 & 7972 & 17010 & - & - & - & - & - & - \\
$\FFTGs(29^2,r^2)$  & 16380 & 4640 & 17455 & 39361 & 16215 & 39544 & 16828 & 7350 & 17334 & - & - & - & - & - & - & - & - & - \\
$\FFTGs(30^2,r^2)$  & 39361 & 12862 & 40316 & 16828 & 5904 & 17696 & - & - & - & - & - & - & - & - & - & - & - & - \\
$\FFTGs(31^2,r^2)$  & 16828 & 4618 & 18062 & - & - & - & - & - & - & - & - & - & - & - & - & - & - & - \\

      \bottomrule
    \end{tabular}
    \label{table:fft3flops}
  \end{table*}

    \begin{table*} \centering
    \small
    \setlength\tabcolsep{2.5pt}
    \caption{AIs of Gauss--FFT transforms for a single tile or kernel.}
    \begin{tabular}{c | ccc | ccc | ccc | ccc | ccc | ccc}
      \toprule
      & \multicolumn{3}{c|}{$r = 2$}& \multicolumn{3}{c|}{$r = 3$}& \multicolumn{3}{c|}{$r = 4$}& \multicolumn{3}{c|}{$r = 5$}& \multicolumn{3}{c|}{$r = 6$}& \multicolumn{3}{c}{$r = 7$} \\
      Method & In & Ker & Out  & In & Ker & Out  & In & Ker & Out  & In & Ker & Out  & In & Ker & Out  & In & Ker & Out \\
      \midrule

      $\FFTGs(2^2,r^2)$  & 0.58 & 0.61 & 0.68 & 0.42 & 0.44 & 0.50 & 0.96 & 1.05 & 1.03 & 0.78 & 0.85 & 0.76 & 1.41 & 1.52 & 1.10 & 0.76 & 0.83 & 0.64 \\
      $\FFTGs(3^2,r^2)$  & 0.42 & 0.38 & 0.54 & 0.96 & 1.02 & 1.09 & 0.78 & 0.75 & 0.83 & 1.41 & 1.52 & 1.46 & 0.76 & 0.81 & 0.76 & 1.59 & 1.70 & 1.46 \\
      $\FFTGs(4^2,r^2)$  & 0.96 & 0.72 & 1.12 & 0.78 & 0.71 & 0.86 & 1.41 & 1.50 & 1.50 & 0.76 & 0.77 & 0.85 & 1.59 & 1.69 & 1.52 & 1.16 & 1.21 & 1.23 \\
      $\FFTGs(5^2,r^2)$  & 0.78 & 0.66 & 0.89 & 1.41 & 1.14 & 1.53 & 0.76 & 0.74 & 0.91 & 1.59 & 1.51 & 1.58 & 1.16 & 1.15 & 1.26 & 2.22 & 2.39 & 2.31 \\
      $\FFTGs(6^2,r^2)$  & 1.41 & 0.77 & 1.53 & 0.76 & 0.65 & 0.93 & 1.59 & 1.30 & 1.60 & 1.16 & 1.12 & 1.29 & 2.22 & 2.37 & 2.35 & 0.98 & 1.01 & 1.18 \\
      $\FFTGs(7^2,r^2)$  & 0.76 & 0.55 & 0.93 & 1.59 & 1.13 & 1.60 & 1.16 & 1.09 & 1.31 & 2.22 & 1.98 & 2.37 & 0.98 & 1.03 & 1.19 & 2.18 & 2.27 & 2.32 \\
      $\FFTGs(8^2,r^2)$  & 1.59 & 0.82 & 1.59 & 1.16 & 1.01 & 1.32 & 2.22 & 1.58 & 2.37 & 0.98 & 1.00 & 1.17 & 2.18 & 2.24 & 2.33 & 1.61 & 1.61 & 1.74 \\
      $\FFTGs(9^2,r^2)$  & 1.16 & 0.73 & 1.32 & 2.22 & 1.18 & 2.36 & 0.98 & 0.90 & 1.16 & 2.18 & 2.22 & 2.40 & 1.61 & 1.58 & 1.75 & 1.55 & 1.65 & 1.67 \\
      $\FFTGs(10^2,r^2)$  & 2.22 & 0.80 & 2.33 & 0.98 & 0.86 & 1.15 & 2.18 & 2.07 & 2.40 & 1.61 & 1.55 & 1.76 & 1.55 & 1.60 & 1.67 & 1.15 & 1.14 & 1.32 \\
      $\FFTGs(11^2,r^2)$  & 0.98 & 0.64 & 1.14 & 2.18 & 1.77 & 2.38 & 1.61 & 1.35 & 1.76 & 1.55 & 1.50 & 1.74 & 1.15 & 1.09 & 1.33 & 2.70 & 2.82 & 2.99 \\
      $\FFTGs(12^2,r^2)$  & 2.18 & 0.96 & 2.36 & 1.61 & 1.13 & 1.75 & 1.55 & 1.38 & 1.74 & 1.15 & 1.03 & 1.33 & 2.70 & 2.67 & 2.99 & 1.85 & 1.75 & 1.78 \\
      $\FFTGs(13^2,r^2)$  & 1.61 & 0.76 & 1.75 & 1.55 & 1.20 & 1.73 & 1.15 & 0.98 & 1.34 & 2.70 & 2.54 & 2.97 & 1.85 & 1.71 & 1.79 & 3.97 & 3.78 & 3.97 \\
      $\FFTGs(14^2,r^2)$  & 1.55 & 0.83 & 1.72 & 1.15 & 0.82 & 1.33 & 2.70 & 2.44 & 2.96 & 1.85 & 1.46 & 1.80 & 3.97 & 3.67 & 3.96 & 1.38 & 1.30 & 1.55 \\
      $\FFTGs(15^2,r^2)$  & 1.15 & 0.66 & 1.33 & 2.70 & 2.03 & 2.93 & 1.85 & 1.17 & 1.79 & 3.97 & 3.37 & 3.93 & 1.38 & 1.21 & 1.55 & 2.01 & 2.19 & 2.10 \\
      $\FFTGs(16^2,r^2)$  & 2.70 & 1.18 & 2.90 & 1.85 & 1.00 & 1.79 & 3.97 & 3.15 & 3.94 & 1.38 & 1.17 & 1.55 & 2.01 & 1.98 & 2.20 & 2.42 & 1.99 & 2.61 \\
      $\FFTGs(17^2,r^2)$  & 1.85 & 0.82 & 1.77 & 3.97 & 3.02 & 3.91 & 1.38 & 1.11 & 1.55 & 2.01 & 1.88 & 2.19 & 2.42 & 1.78 & 2.60 & 5.06 & 4.74 & 5.43 \\
      $\FFTGs(18^2,r^2)$  & 3.97 & 2.28 & 3.86 & 1.38 & 1.02 & 1.55 & 2.01 & 1.59 & 2.18 & 2.42 & 1.60 & 2.60 & 5.06 & 4.67 & 5.40 & 1.34 & 1.20 & 1.54 \\
      $\FFTGs(19^2,r^2)$  & 1.38 & 0.75 & 1.54 & 2.01 & 1.22 & 2.17 & 2.42 & 1.36 & 2.58 & 5.06 & 3.77 & 5.36 & 1.34 & 1.18 & 1.54 & 2.79 & 2.05 & 2.61 \\
      $\FFTGs(20^2,r^2)$  & 2.01 & 0.84 & 2.16 & 2.42 & 1.15 & 2.57 & 5.06 & 3.54 & 5.31 & 1.34 & 1.11 & 1.52 & 2.79 & 1.90 & 2.60 & 2.38 & 2.10 & 2.60 \\
      $\FFTGs(21^2,r^2)$  & 2.42 & 0.79 & 2.55 & 5.06 & 3.30 & 5.26 & 1.34 & 0.99 & 1.52 & 2.79 & 1.76 & 2.59 & 2.38 & 2.06 & 2.59 & 2.92 & 2.06 & 2.64 \\
      $\FFTGs(22^2,r^2)$  & 5.06 & 2.84 & 5.27 & 1.34 & 0.88 & 1.50 & 2.79 & 1.51 & 2.58 & 2.38 & 1.90 & 2.59 & 2.92 & 1.93 & 2.63 & 1.83 & 1.62 & 2.01 \\
      $\FFTGs(23^2,r^2)$  & 1.34 & 0.69 & 1.49 & 2.79 & 1.28 & 2.56 & 2.38 & 1.78 & 2.57 & 2.92 & 1.68 & 2.62 & 1.83 & 1.57 & 2.00 & 4.15 & 3.76 & 4.46 \\
      $\FFTGs(24^2,r^2)$  & 2.79 & 0.85 & 2.54 & 2.38 & 1.72 & 2.56 & 2.92 & 1.43 & 2.60 & 1.83 & 1.51 & 2.00 & 4.15 & 3.67 & 4.44 & 1.75 & 1.53 & 1.95 \\
      $\FFTGs(25^2,r^2)$  & 2.38 & 0.90 & 2.54 & 2.92 & 1.22 & 2.59 & 1.83 & 1.41 & 1.99 & 4.15 & 3.58 & 4.41 & 1.75 & 1.44 & 1.95 & 4.02 & 3.48 & 4.36 \\
      $\FFTGs(26^2,r^2)$  & 2.92 & 0.87 & 2.57 & 1.83 & 1.11 & 1.98 & 4.15 & 3.25 & 4.38 & 1.75 & 1.37 & 1.94 & 4.02 & 3.46 & 4.33 & 1.58 & 1.38 & 1.78 \\
      $\FFTGs(27^2,r^2)$  & 1.83 & 0.78 & 1.97 & 4.15 & 3.11 & 4.34 & 1.75 & 1.14 & 1.93 & 4.02 & 3.34 & 4.31 & 1.58 & 1.29 & 1.77 & - & - & - \\
      $\FFTGs(28^2,r^2)$  & 4.15 & 2.47 & 4.34 & 1.75 & 0.97 & 1.92 & 4.02 & 3.24 & 4.28 & 1.58 & 1.20 & 1.76 & - & - & - & - & - & - \\
      $\FFTGs(29^2,r^2)$  & 1.75 & 0.80 & 1.91 & 4.02 & 2.71 & 4.24 & 1.58 & 1.11 & 1.75 & - & - & - & - & - & - & - & - & - \\
      $\FFTGs(30^2,r^2)$  & 4.02 & 2.16 & 4.22 & 1.58 & 0.90 & 1.75 & - & - & - & - & - & - & - & - & - & - & - & - \\
      $\FFTGs(31^2,r^2)$  & 1.58 & 0.71 & 1.74 & - & - & - & - & - & - & - & - & - & - & - & - & - & - & - \\

      \bottomrule
    \end{tabular}
    \label{table:fft3ai}
  \end{table*}

  In Tbl.~\ref{table:winflops} and Tbl.~\ref{table:winai} we show the lookup tables used in our model  to determine the number of required operations, and the arithmetic intensities for transforming
 a single input/output tile, or a single kernel for the Winograd method.
  
 In Tbl.~\ref{table:fftflops} and Tbl.~\ref{table:fftai} we show the lookup tables used in our model
  to determine the number of required operations, and the arithmetic intensities for transforming
  a single input/output tile, or a single kernel for the Regular--FFT method.
  
  In Tbl.~\ref{table:fft3flops} and Tbl.~\ref{table:fft3ai} we show the lookup tables used in our model
  to determine the number of required operations, and the arithmetic intensities for transforming
 a single input/output tile, or a single kernel for the Gauss--FFT method.

 \section{Additional Model Verification}

%   \begin{figure*}[ht]
%     \begin{center}
%       \includegraphics[width=1\linewidth]{__fig/vs_8.pdf}
%     \end{center}
%     \caption{Theoretical estimates of our model, and empirical
%       measurements for the speedup of Regular-- and Gauss--FFT methods
%       over the Winograd method with lower precision (tile sizes allowed to be up to $8 \times 8$), on VGG and AlexNet.}
%     \label{fig:model2}
%   \end{figure*}
  
  \begin{figure*}[ht]
    \begin{center}
      \includegraphics[width=1\linewidth]{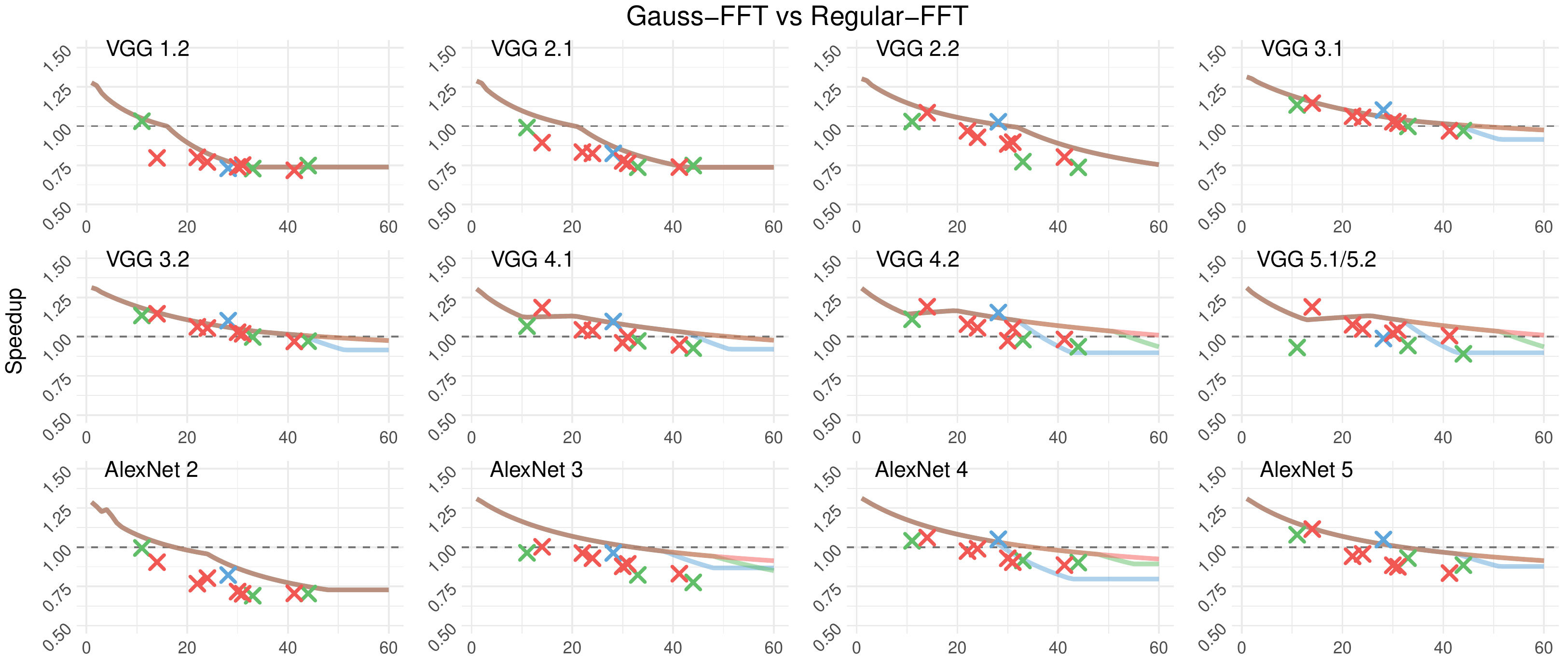}
    \end{center}
    \caption{Theoretical estimates of our model, and empirical
      measurements for the speedup of the Regular-- vs the Gauss--FFT method
      VGG and AlexNet.}
    \label{fig:model3}
  \end{figure*}

%   On Fig.~\ref{fig:model2} we show the theoretical predictions of our model, and the empirical results
%   for the relative performances of Regular-- and Gauss--FFT methods versus the Winograd method with low accuracy (
%   with the tile sizes allowed to be up to $8 \times 8$.
  
 In Fig.~\ref{fig:model3} we show the theoretical predictions and the empirical results of the Gauss--FFT vs the Regular--FFT methods.
  
 \section{Comparing to State--of--the--art}

  \begin{figure*}[ht]
    \begin{center}
      \includegraphics[width=0.95\linewidth]{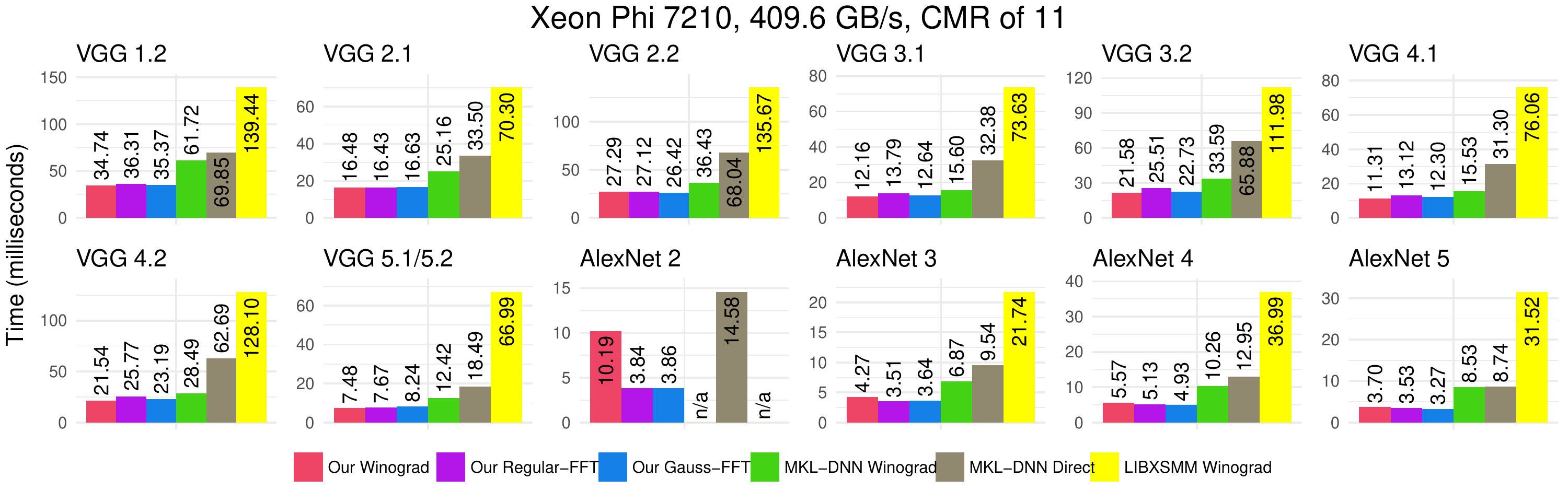}\\
      \includegraphics[width=0.95\linewidth]{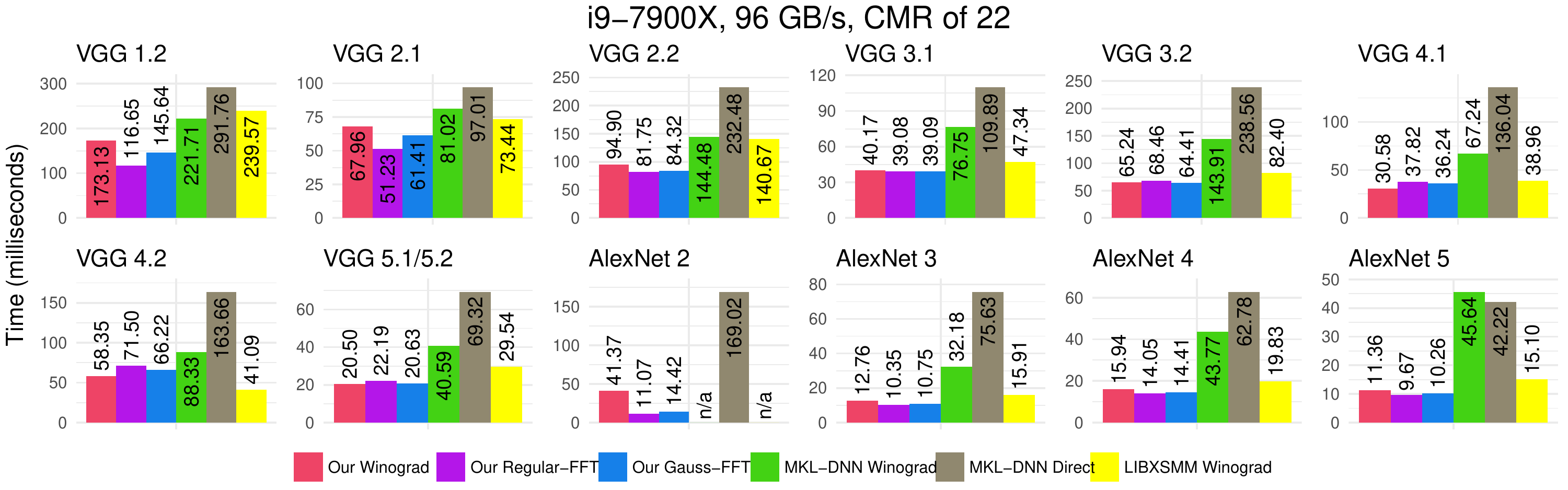}\\
      \includegraphics[width=0.95\linewidth]{__fig/__benchs_xeonsp.pdf}
      \includegraphics[width=0.95\linewidth]{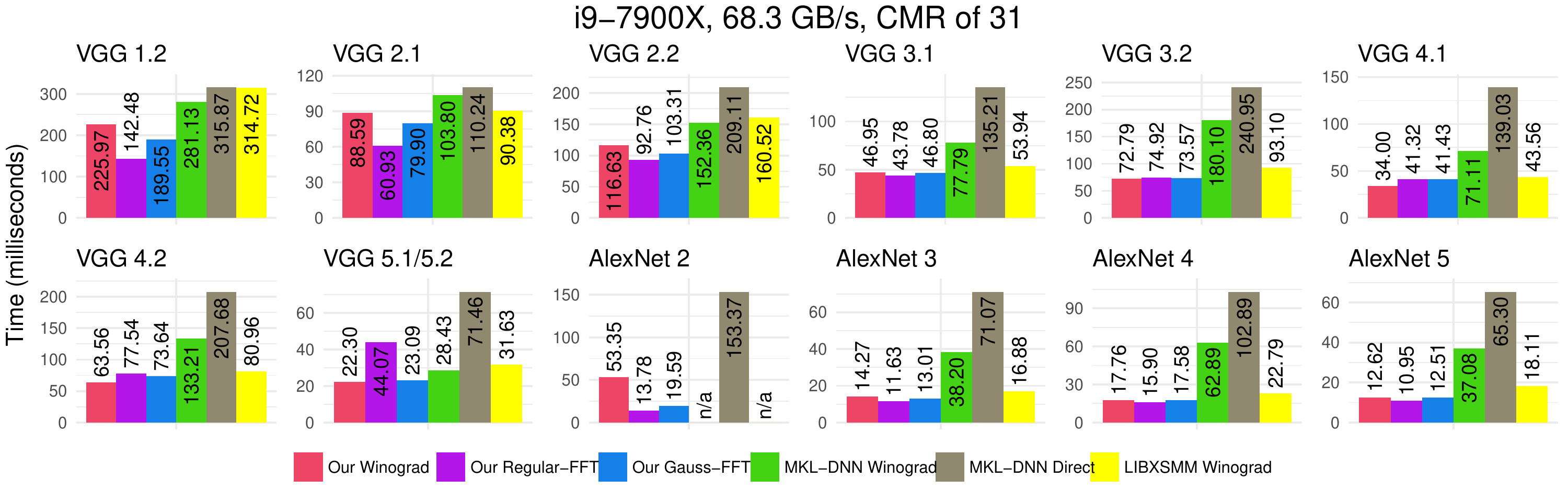}
    \end{center}
    \caption{Running time of our implementations compared to state--of--the--art, on AVX512 capable systems.}
    \label{fig:others2}
  \end{figure*}

  \begin{figure*}[ht]
    \begin{center}
      \includegraphics[width=0.95\linewidth]{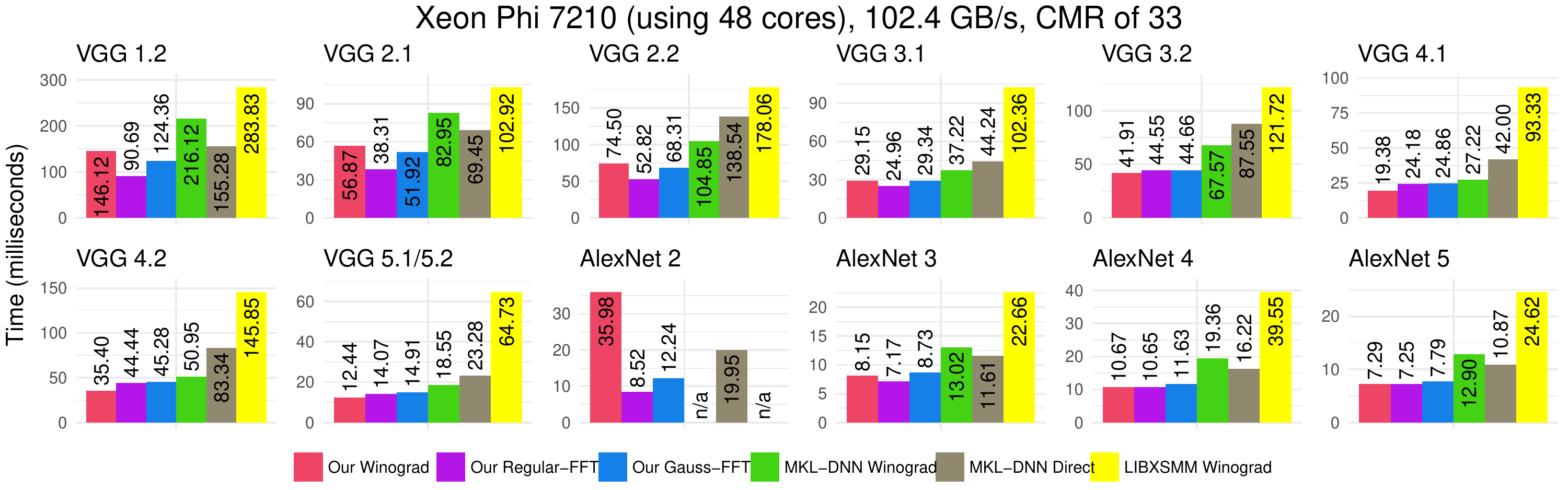}
      \includegraphics[width=0.95\linewidth]{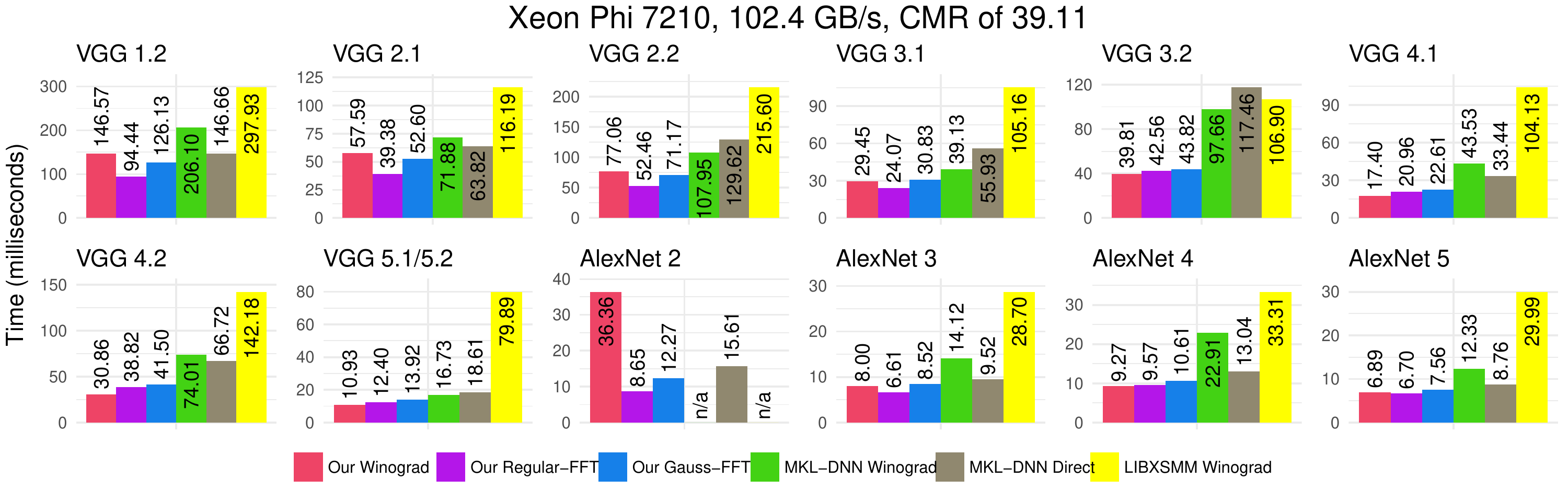}
      \includegraphics[width=0.95\linewidth]{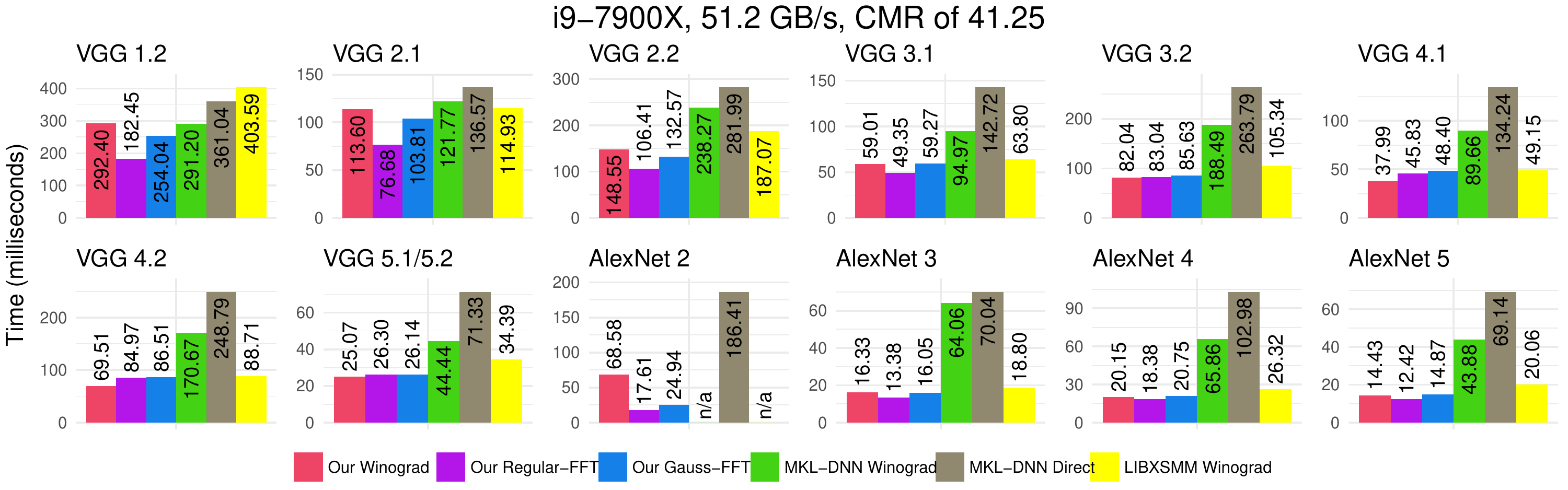}
    \end{center}
    \caption{Running time of our implementations compared to state--of--the--art on AVX512 capable systems, continued.}
    \label{fig:others3}
  \end{figure*}

   In Fig.~\ref{fig:others2} and Fig.~\ref{fig:others3} we show the results of the absolute time
  required for processing unique layers of VGG and AlexNet of our implementation and other, state--of--the--art
   libraries.  The benchmarks were performed on AVX512 capable systems, as the neither MKL-DNN nor LIBXSMM support AVX2.
  
   Both MKL-DNN and LIBXSMM support only kernels of size $3 \times 3$ for their Winograd implementation; thus, the 
  measurements for the layer 2 of AlexNet, which have kerenls of size $5 \times 5$ are missing.

\end{document}